\documentclass[12pt]{article}
\usepackage[usenames, dvipsnames]{xcolor}
\usepackage{jcapmod}

\usepackage{tocloft}
\usepackage[normalem]{ulem}
\usepackage[]{todonotes}
\usepackage{verbatim}
\usepackage{mathrsfs}
\usepackage{cleveref}
\usepackage[margin=.8in]{geometry}

\usepackage{tikz}
\usepackage{setspace,caption}
\usepackage{lipsum}
\usetikzlibrary{matrix,arrows,calc}
\usepackage{bbm}          
\usepackage{slashed}        
\usepackage{graphicx}    
\usepackage{subcaption} 
\usepackage{psfrag}    
\usepackage{tensor}      
\usepackage{fouridx}     
\usepackage{bm}         
\usepackage{mdframed}      
\usepackage{multirow}  
\usepackage{soul}  
\usepackage{bbold}  
\usepackage{multicol} 
\usepackage{tikz-cd}
\usepackage{rotating}

\usetikzlibrary{arrows}

  \usepackage{mdframed} 
\tikzset{
commutative diagrams/.cd,
arrow style=tikz,
diagrams={>=latex}}

\newcommand{\aside}[2]{\vspace{4pt} \begin{mdframed} #1 #2 \end{mdframed}\vspace{4pt}}

\usepackage{amsmath}
\usepackage{amssymb}
\usepackage{mathtools}

\usepackage{feynmf}
\usepackage{marvosym}

\usepackage{import}

\usepackage[utf8]{inputenc}
\usepackage{amsmath}
\usepackage{tikz}
\usepackage{extpfeil}
\usepackage{amsfonts}
\usepackage{amssymb} 
\usepackage{bm}
\usepackage{graphicx}
\usepackage{array}
\usepackage{lipsum}
\usepackage{float}
\usepackage{multirow}
\usepackage{hhline}
\usepackage{tabularx}
\usepackage{graphicx}
\usepackage{afterpage}
\usepackage{longtable}
\usepackage{epstopdf}

\usepackage[titletoc]{appendix}

\newcommand{\cO}{\mathcal{O}}

\newcommand{\bZ}{\mathbb{Z}}
\newcommand{\bC}{\mathbb{C}}

\newcommand{\jht}[1]{{}}
\newcommand{\jt}[1]{{}}

\definecolor{cobalt}{RGB}{44, 98, 120}
\definecolor{celadon}{rgb}{0.67, 0.88, 0.69}
\definecolor{dm}{cmyk}{.20, 0, .30, 0}
\definecolor{burgundy}{rgb}{0.5, 0.0, 0.13}
\definecolor{plotBlue}{RGB}{94, 130, 181}

\newcommand*\xoverline[2][0.75]{
    \sbox{\myboxA}{$\m@th#2$}
    \setbox\myboxB\null
    \ht\myboxB=\ht\myboxA
    \dp\myboxB=\dp\myboxA
    \wd\myboxB=#1\wd\myboxA
    \sbox\myboxB{$\m@th\overline{\copy\myboxB}$}
    \setlength\mylenA{\the\wd\myboxA}
    \addtolength\mylenA{-\the\wd\myboxB}
    \ifdim\wd\myboxB<\wd\myboxA
       \rlap{\hskip 0.5\mylenA\usebox\myboxB}{\usebox\myboxA}%
    \else
        \hskip -0.5\mylenA\rlap{\usebox\myboxA}{\hskip 0.5\mylenA\usebox\myboxB}%
    \fi}
\makeatother

\begin{document}

\newcommand{\main}{.}
\begin{titlepage}

\setcounter{page}{1} \baselineskip=15.5pt \thispagestyle{empty}

\bigskip\

\vspace{2cm}
\begin{center}
{\LARGE \bfseries Non-simply-laced Symmetry Algebras\\ \vspace{.2cm} in F-theory on Singular Spaces}
\end{center}

\vspace{0.25cm}

\begin{center}
Antonella Grassi$^1$,
James Halverson$^2$, Cody Long$^2$, Julius L. Shaneson$^1$,  and Jiahua Tian$^2$ \\ 

\vspace{1 cm}
\emph{$^1$Department of Mathematics, University of Pennsylvania \\ Philadelphia, PA 19104}\\
\vspace{.3cm}
\emph{$^2$Department of Physics, Northeastern University \\ Boston, MA 02115}\\

\vspace{0.5cm}
\end{center}

\vspace{1cm}
\noindent

We demonstrate how non-simply-laced gauge and flavor symmetries arise in F-theory
on spaces with non-isolated singularities. The breaking from a simply-laced symmetry to one that is non-simply-laced
is induced by Calabi-Yau complex structure deformation. In all examples the deformation maintains non-isolated
singularities but is accompanied
by a splitting of an $I_1$ seven-brane that opens new loops in the geometry near a 
non-abelian seven-brane. The splitting also arises in the moduli
space of a probe D3-brane, which
upon traversing the new loop experiences a monodromy that acts on 3-7 string junctions on the
singular space.
The monodromy reduces the symmetry algebra, which is the flavor symmetry of the
D3-brane and the gauge symmetry of the seven-brane, to one that is non-simply-laced. A collision of the D3-brane with the seven-brane gives
rise to a 4d $\mathcal{N}=1$ SCFT with a non-simply-laced flavor symmetry.

\noindent
 \vspace{3.1cm}

\end{titlepage}
\tableofcontents
\newpage

\section{Introduction}

F-theory~\cite{Vafa:1996xn, Morrison:1996pp} is a strongly coupled generalization of type IIb string theory that allows for a varying axio-dilaton in the internal space, including regions where the string coupling constant is $O(1)$. The power of F-theory lies in the fact that it geometrizes 7-branes, by promoting the axiodilaton to be the complex structure of an elliptic curve fibered over the physical internal space $B$, so that seven-branes in F-theory are described by a Calabi-Yau elliptic (or genus-one) fibration $X \rightarrow B$. The singularities of the elliptic fibration encode the positions and types of 7-branes, and provides the fundamental geometric data to compute the gauge group, matter, and other physical content of the low-energy effective theory.

Understanding an F-theory compactification is often done by smoothing the elliptic Calabi-Yau $X$ in some fashion. One well-studied approach is to resolve the singularities associated with seven-brane in $X$ via a series of blowups and small resolutions~\cite{Bershadsky:1996nh,Katz:1996th, Morrison:2011mb, Katz:2011qp, Esole:2011sm, Marsano:2011hv, Lawrie:2012gg, Hayashi:2014kca, Klevers:2014bqa} to obtain a smooth Calabi-Yau fourfold $X^\sharp$. Compactification of M-theory on $X^\sharp$ corresponds to the Coulomb branch associated with a seven-brane gauge theory. However, this Coulomb branch arises only via a circle compactification of the original theory; it does not exist in the F-theory limit and, though very useful, it is nevertheless indirect. Another approach is to deform~\cite{Gaberdiel:1997ud, DeWolfe:1998bi, DeWolfe:1998zf, Grassi:2013kha,Grassi:2014sda,Grassi:2014zxa,Grassi:2014ffa,Grassi:2016bhs} the complex structure of $X$ to obtain a smooth Calabi-Yau $X^\flat$, which typically corresponds to a rank-reducing Higgsing of the gauge group. This has the advantage that this branch of the moduli space exists in the F-theory limit, and also in the associated M-theory compactification in one dimension less.

\vspace{.5cm}
However, from the point of view of gauge theory this seems rather strange:  why must the gauge group be broken in order to understand the unbroken theory? Of course, it need
not: it is only a matter
of mathematical and technical convenience, and in general doing so will miss
some of the physics of the unbroken theory. More specifically, the mathematical
techniques for studying F-theory on singular spaces $X$ are simply not as well-developed as on
its smoothings, and dimensional reduction of M-theory on a singular space
$X$ is not well understood in general\footnote{One case that is well understood are
those related 
to weakly-coupled IIb orientifolds with non-abelian D7-brane configurations: the corresponding elliptically-fibered Calabi-Yau $X$ is still singular, but we understand the theory well due to the associated string theory, and not the M-theory compactification on the singular space. }.

We will therefore study F-theory directly in cases where $X$ has non-isolated singularities, which corresponds to
having a non-abelian gauge group $G$ on seven-branes. 
Progress in this direction,
rather than studying F-theory on a smoothing of $X$ that breaks $G$, seems critical
for a number of reasons:
\begin{itemize}
\item \textbf{Naturalness.} It seems much more natural to study a gauge theory directly,
rather than via its broken phases. The unbroken phase arises for singular $X$. 
\item \textbf{Moduli space obstruction.} The singular theory may exist at the intersection of
multiple branches of moduli space, and by moving to the deformation or resolution 
some other branches may be obstructed, for example those corresponding to T-branes~\cite{Cecotti:2010bp}. 

\item \textbf{Calabi-Yau smoothing is often impossible.} There is increasing
evidence that typical $X$ have both non-Higgsable clusters (see, e.g., \cite{Halverson:2015jua,Taylor:2015ppa,Halverson:2017ffz,Taylor:2017yqr}) and terminal singularities~\cite{Arras:2016evy},
which forbid passing to smooth Calabi-Yau varieties by complex structure deformation
and K\"ahler  resolution, respectively.
\end{itemize}
Some excellent progress has already been made in this direction \cite{Collinucci:2014taa}. However, it also seems that there are many essential questions in F-theory that have yet
to be answered from the point of view of the singular geometry that are intrinsic to 
its non-abelian gauge sectors.

\vspace{.5cm}
In this paper we present a conceptually clean result that derives well-known F-theoretic
phenomena, but in F-theory on a singular space.\footnote{By this we mean that the
elliptic fibration $X$ has non-isolated singularities and has non-abelian seven-branes, but its base $B$,
which make up the extra dimensions of space, is
smooth.} Our tool will be string junctions
that begin on a probe D3-brane and end on 7-branes in singular F-theory geometries, which
we motivate using a result from the math literature. This
builds critically on the theory of topological string
junctions~\cite{Grassi:2013kha, Grassi:2014sda,Grassi:2014zxa,Grassi:2014ffa,Grassi:2016bhs,Grassi:toappear}.
Specifically, we will 
derive the existence of non-simply-laced gauge groups on seven-branes,
which correspond to non-simply-laced flavor symmetries on the D3-brane, from monodromy actions on the 3-7 string junctions.

The existence of such non-simply-laced symmetry groups, including the exceptional examples $F_4$
and $G_2$,
is a classic result in F-theory. It is well understood from the perspective of the
smooth resolution $X^\sharp$, where monodromy action on the
generic Kodaira fiber dictates the non-simply-laced structure. This monodromy action 
begins to act when the non-abelian fiber transitions from split to semi-split or non-split,
which corresponds to Higgsing from a simply-laced group to a non-simply-laced group;
in both cases the analysis is done by passing to the M-theory Coulomb branch \cite{Morrison:1996pp,Bershadsky:1996nh} in one dimension less. 
Similarly, in complex structure deformations to a smooth Calabi-Yau $X^\flat$,
corresponding to the Higgs branch,
monodromy action on string junctions can give rise to non-simply-laced groups~\cite{Bonora:2010bu,Grassi:2013kha}.

Our analysis on singular spaces relies heavily on one critical observation: in deforming from
\begin{equation}
X_{sl}\longrightarrow X_{nsl},
\end{equation}
i.e. from the singular space associated with a simply-laced group to that of a non-simply-laced group, new non-trivial loops in the D3 moduli space appear, due to a generic splitting of $I_1$ loci\footnote{Such deformations are from specialized loci to generic regions in the moduli space that preserves the Kodaira fiber.}. It is precisely the monodromy action associated to these loops that will reduce the gauge algebra from simply-laced to non-simply-laced, via an action on the charge and representations of the strings. This occurs in every example that we study and presents a new technique for understanding 4d $\mathcal{N}=1$ SCFTs with non-simply-laced flavor symmetries, such as those studied in \cite{Aharony:1996bi}. Since this observation arises only from Higgsing a simply-laced group to a non-simply-laced group, without resolving or deforming to a smooth manifold, it could also be understood as the origin of non-simply-laced groups in the F-theory limit.

We emphasize at the outset that there are two different notions of ``splitting" that we will use. One is the splitting in Kodaira's sense, determined by whether there is an outer-automorphism of the Dynkin diagram corresponding to the fibration structure. The other is the splitting associated to the creation of the new closed loops in the D3 brane moduli space, which we will make precise below. When a Kodaira fiber becomes non-split (or semi-split), the $I_1$ locus becomes split near the seven-brane, and a simply-laced group is broken to a non-simply-laced group. The two notions of splitting should be clear from context.

This paper is organized as follows. In section \ref{sec:geom_monodromy} we 
analytically compute various geometric monodromies, including those associated with
$I_1$ locus splitting, using a simple technique. In section \ref{sec:junc_analysis} we review string junctions
on deformed spaces, discuss aspects of them on singular spaces, and how monodromy
reduction to non-simply laced algebras occurs on singular spaces.

\section{Analytic Computation of Geometric Monodromies}\label{sec:geom_monodromy}

The crux of our analysis is that under a deformation from split non-abelian fiber to non-split non-abelian fiber over a divisor $D$, the intersection of the $I_1$ locus with $D$ can split, thereby providing new non-trivial loops in the vicinity
of the seven-brane.  These new paths can provide new monodromy actions on 3-7 string junctions as a D3-brane traverses the loop, and we will find that such monodromy actions reduce the symmetry algebra to one that is non-simply-laced.

Our analysis proceeds in two steps: we first compute the relevant
monodromies around irreducible $I_1$ loci that arise in the deformation from split non-abelian fiber to non-split non-abelian fiber, and then compute the induced monodromy action on representations of the
gauge algebra, which arise from string junctions ending on the non-abelian seven-branes.
In this section, we first explain the $I_1$ splitting phenomenon as we move from split non-abelian fiber to non-split non-abelian fiber, and then
present a simple technique for analytically determining the
vanishing cycles associated to seven-branes with Kodaira $I_1$ fibers. We will end
this section with the computation of geometric monodromies in a number of examples. 

\subsection{The central observation: splitting $I_1$ loci}\label{sec:gen6D}

An elliptic curve can be regarded as a double cover of $\mathbb{P}^1$ with four punctures
at which the double cover is ramified. In
a Weierstrass model, which takes the form 
\begin{equation}y^2 = x^3 + fx + g\, ,
\end{equation}
three of the punctures are manifest as the roots of the cubic $x^3 + fx + g = 0$, denoted by $x_1$, $x_2$ and $x_3$,
while the fourth root lies at infinity. In an elliptic fibration over a base $B$,
\begin{equation}
X \xrightarrow{\pi} B\, ,
\end{equation} 
the positions of the three punctures vary as we move around on the base manifold $B$ of the fibration. The requirement that $X$ is Calabi-Yau then implies that $f\in \Gamma(-4K_B)$
and $g\in \Gamma(-6K_B)$, where $-K_B$ is the anticanonical bundle of $B$.

At a point $u_D$ on the discriminant locus 
 \begin{equation}
 \Delta=4f^3+27g^2=0\, ,
 \end{equation}
 the elliptic curve degenerates, i.e. $\pi^{-1}(u_D)$ is a singular fiber,
where at least two of $\{x_1, x_2, x_3\}$ have collided. If we move to a point $p$, slightly away from the discriminant locus, and take a loop around a component of the discriminant locus, this will induce a non-trivial map on $\{x_1, x_2, x_3\}$, which in turn induces a monodromy action on $H_1(E_p,\mathbb{Z})$, where 
\begin{equation}
E_p:=\pi^{-1}(p)\, ,
\end{equation}
 is the smooth elliptic curve above $p$. If the singular fiber above the discriminant component is of Kodaira type $I_1$, then only two of the roots of the cubic become degenerate upon approaching this $I_1$ component. If we take $p$ sufficiently close to such a locus, two of the roots will be nearly collided, and upon encircling the nearby $I_1$ locus those two roots swap. This swap precisely determines the Picard-Lefschetz monodromy on $H_1(E_p,\mathbb{Z})$ associated with traversing the loop.

Now note that, as a function of $f$, there are three solutions to the equation $\Delta\equiv 4f^3 + 27g^2 = 0$, when restricted to a local patch of the base manifold $B$ (where $f$ and $g$ can be treated locally as ordinary functions on $B$). These roots take the form:
\begin{align}
  f_1 &= -3 \left(-\frac{1}{2}\right)^{2/3} g^{2/3}\, , \label{eq:froot1} \\
  f_2 &= -\frac{3 g^{2/3}}{2^{2/3}}\, , \label{eq:froot2} \\
  f_3 &= \frac{3 \sqrt[3]{-1} g^{2/3}}{2^{2/3}}\, , \label{eq:froot3}
\end{align}
where here and henceforth when we indicate a cube root, we mean the principal cube root, i.e. the one with least non-negative argument. Upon traversing a loop around an  $I_1$-component of the discriminant locus two of the roots $\{x_1, x_2, x_3\}$ are swapped, where the particular choice of roots is determined by which one of the three solutions to $\Delta = 0$ (as a function of $f$) is realized
at a given point $u_D\in \{\Delta = 0\}$. Which two in $\{x_1, x_2, x_3\}$ swap
also determines the vanishing cycle.

\begin{figure}[th]
  \centering
  \includegraphics[width=.5\linewidth]{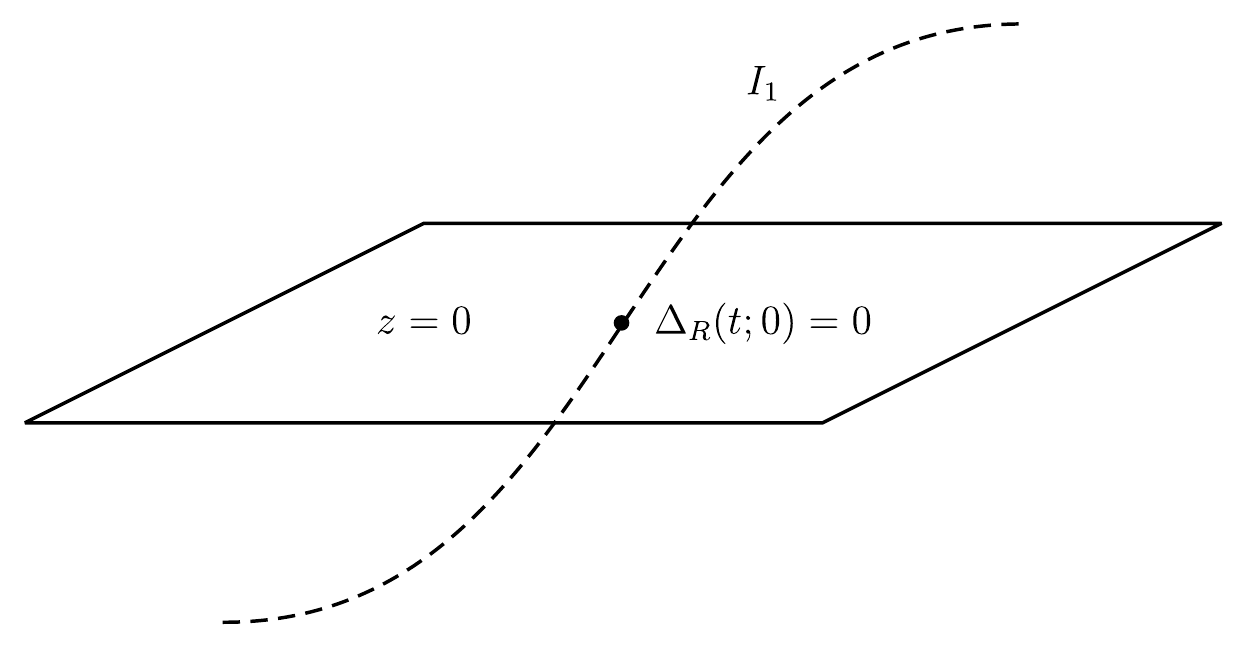}
    \caption{An $I_1$ locus intersects the stack of 7-branes at $z = 0$.}
  \label{fig:7brane+I1}
\end{figure}

Non-simply-laced gauge groups can only be realized in F-theory when $dim_\bC(B)\geq 2$, i.e., in six-dimensional
compactification or lower, and therefore in any local patch the Weierstrass model depends
on multiple complex coordinates. To engineer a gauge group we will consider a singular Weierstrass model with a non-abelian seven-brane at $z=0$, for local coordinates $\{z, t_i\}$ on the base
\begin{equation}\label{eq:genWei}
  \begin{aligned}
    f &= f(t_i; z), \\
    g &= g(t_i; z).
  \end{aligned}
\end{equation}
The discriminant locus then must take the form
\begin{equation}\label{eq:dlocus}
  \Delta = z^N \Delta_R(t_i; z)\, ,
\end{equation}
 where $\Delta_R(t_i; z)=0$ is called the residual discriminant locus, whose generic fiber is of $I_1$-type.\footnote{In some cases, such as those with non-Higgsable clusters, there can be additional non-$I_1$ factors, in which case we still denote the $I_1$-locus
 as $\Delta_R(t;z)$. For any model with $dim_\bC(B) > 2$ there will be multiple $t_i$, but for our analysis it will be sufficient to consider $D$ as a small disk transverse to $I_1$ loci, parameterized by a single complex coordinate $t$.}
The fiber generically becomes more singular along intersections of $z = 0$ and $\Delta_R = 0$, as depicted in Figure~\ref{fig:7brane+I1}. 

\begin{figure}[th]
  \begin{subfigure}{0.5\textwidth}
      \centering
      \includegraphics[width=.7\linewidth]{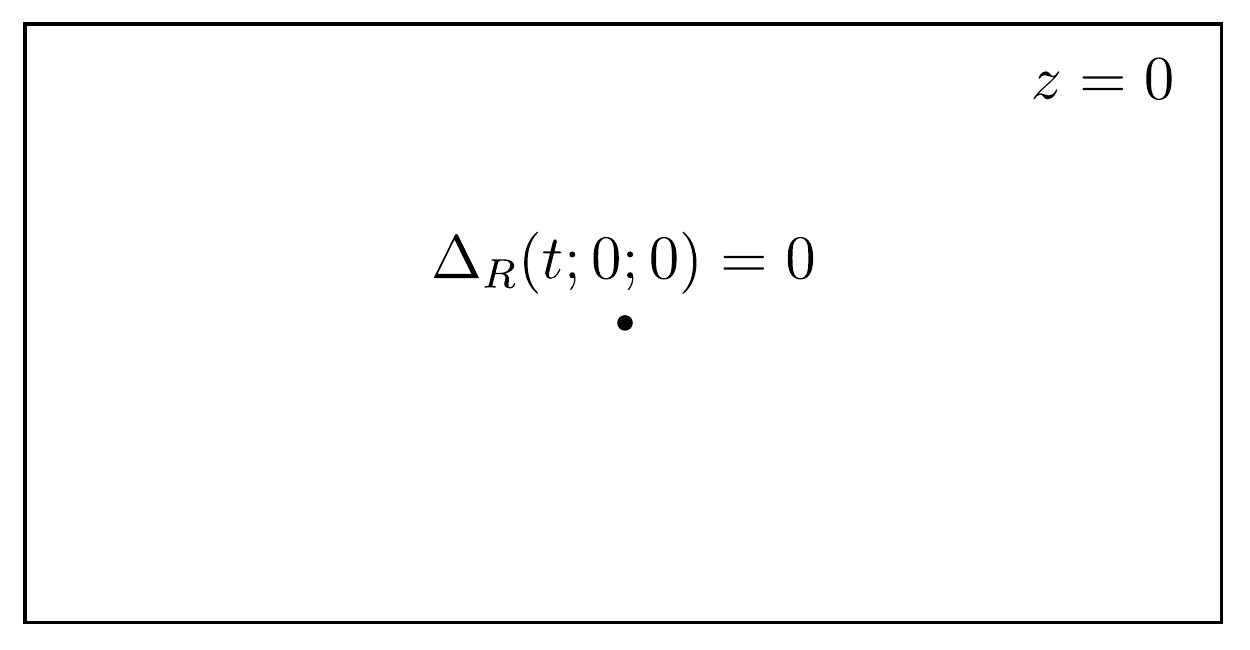}
      \caption{$\epsilon = 0$}
      \label{fig:base_e_off}
  \end{subfigure}%
  $\xrightarrow{\,\,\,\epsilon\neq 0 \,\,\,}$
  \begin{subfigure}{.5\textwidth}
      \centering
      \includegraphics[width=.7\linewidth]{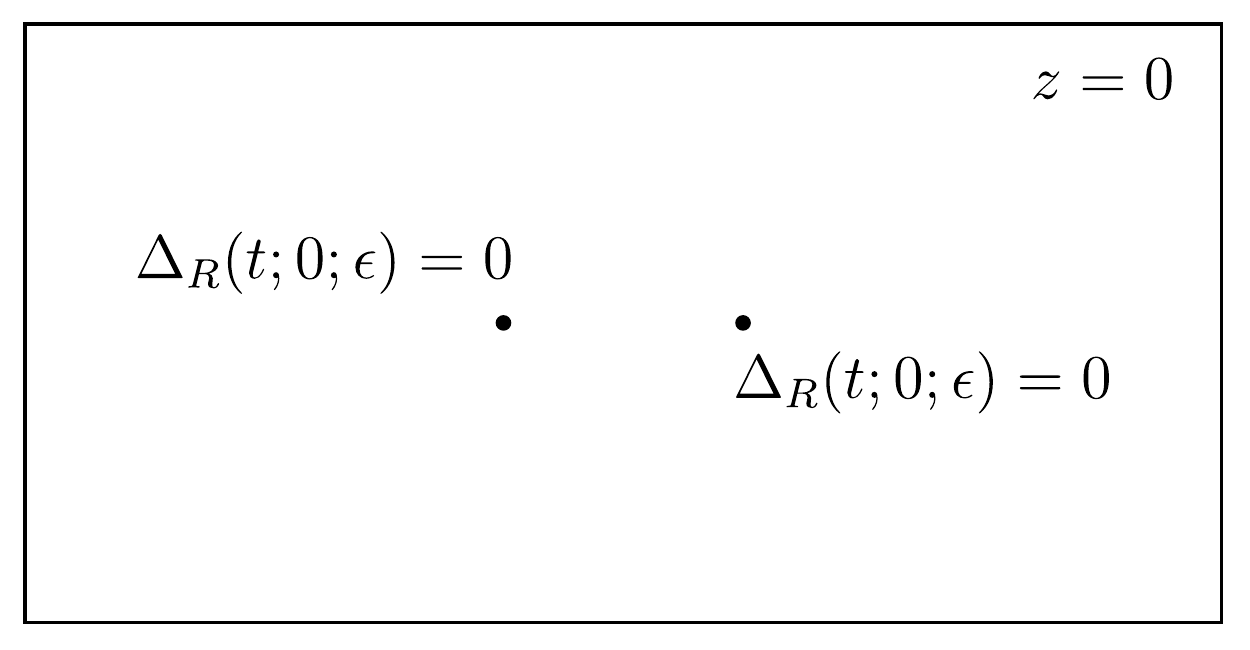}
      \caption{$\epsilon \neq 0$}
      \label{fig:base_e_on}
  \end{subfigure}
  \caption{Splitting of the $I_1$ locus associated with the deformation
    to $\epsilon \neq 0$ that transitions the simply-laced gauge algebra to the 
    non-simply-laced one.}
  \label{fig:base_e}
\end{figure}

Whether or not the gauge group is simply-laced depends on whether or not the associated Kodaira fiber is split. For a detailed explanation of this phenomenon see~\cite{Bershadsky:1996nh} (though we will realize the same phenomenon from different techniques). In each model we study we consider a single-parameter family of deformations, parameterized by an appropriate parameter $\epsilon$, which interpolates between split and non-split Kodaira fibers. To distinguish between these cases we will use the notation $\Delta_R(t; z; \epsilon)$ for the residual discriminant. We choose the parameter $\epsilon$ such that when $\epsilon = 0$ there is no outer-automorphism on the fiber, and therefore when $\epsilon = 0$ the fiber type is split and the gauge algebra is simply-laced, and when $\epsilon \neq 0$ it is non-simply-laced.

On the other hand, when $\epsilon \neq 0 $ and the fiber becomes non-split, each component on the $z = 0$ hyperplane such that $\Delta_R(t; 0; 0) = 0$ can split (and at least one does) into multiple components on the $z = 0$ plane, which are a set of solutions to $\Delta_R(t; 0; \epsilon) = 0$. In this sense the the split vs. non-split issue for the $I_1$ loci has opposite meaning from the non-abelian 7-brane loci, as turning on $\epsilon \neq 0$ such that the non-abelian fiber becomes non-split has the effecting of splitting the $I_1$ loci into multiple components. This scenario is schematically shown in Figure~\ref{fig:base_e}. This is the critical observation for our analysis, and so we place it in a little box:
\aside{A deformation $\epsilon \neq 0$ that changes the non-abelian 7-brane fiber from split to non-split can split the $I_1$ loci intersections with the 7-brane into multiple components. This splitting provides new loops in $D$, and traversing these loops gives the corresponding monodromy action that reduces the
symmetry algebra.}
 
\begin{figure*}[th]
  \centering
  \includegraphics[width=.5\linewidth]{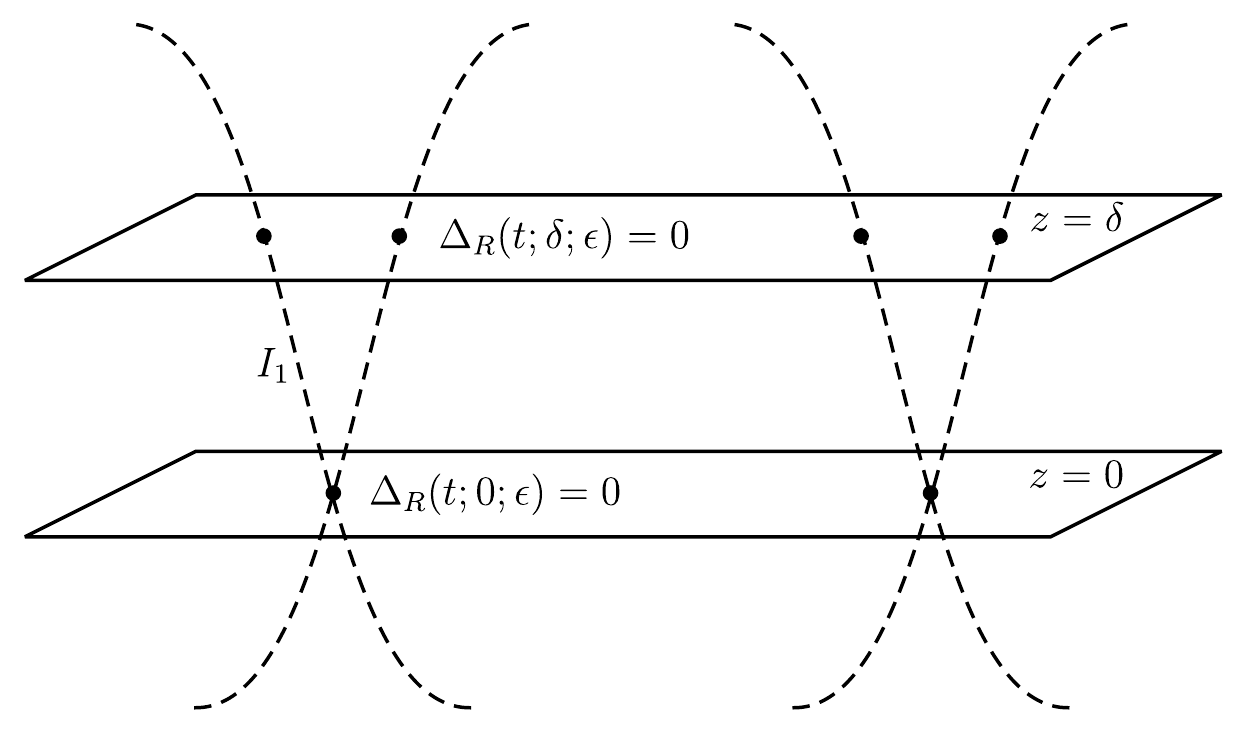}
    \caption{For $IV_{ns}$ and $IV^*_{ns}$, upon turning on $\delta$, the residual discriminant intersection further splits into 2 components along the $z = \delta$ plane. The monodromy
    induced by a path around both both, which coalesces into the monodromy around a single component in the $z=0$
    plane, is the one relevant for reducing to the gauge algebra.}
  \label{fig:schematic_split}
\end{figure*}

The fiber is singular along the non-abelian 7-brane locus $D$, given by $z = 0$, which makes directly probing $D$ difficult. We will bypass this issue by focusing on a 
nearby hyperplane $z = \delta$, $\delta \in \bC$, with $|\delta|$ infinitesimal, in a sense we make precise below. In the examples that we study the non-trivial behavior of taking $z = \delta \neq 0$ is the possibility of further splitting of the components of $\{z = 0\} \cap \{ \Delta_R = 0\}$. For type $I_0^*$ and $I_1^*$, each of the marked points for $\epsilon \neq 0$ remains separated but does not further split. For type $I_4$, $IV$, $IV^*$, each of the components further splits into either three (for $I_4$) or two (for $IV$ and $IV^*$) components on the $z = \delta$ plane. The behavior for $IV_{ns}$ and $IV^*_{ns}$ is shown schematically in Figure~\ref{fig:schematic_split}. We would like to emphasize that the splitting that arises from moving away from the $z = 0$ plane is not the splitting we are interested in, as it does not exist on the non-abelian 7-brane itself. On the other hand, the splitting that occurs by taking $\epsilon \neq 0$ does exist on the non-abelian 7-brane, and such a splitting will be our focus. We therefore will consider loops that encircle all components of an $I_1$ fiber which collapse to a single component on the $z= 0$ hyperplane.

\begin{figure*}[th]
  \centering
  \includegraphics[width=.5\linewidth]{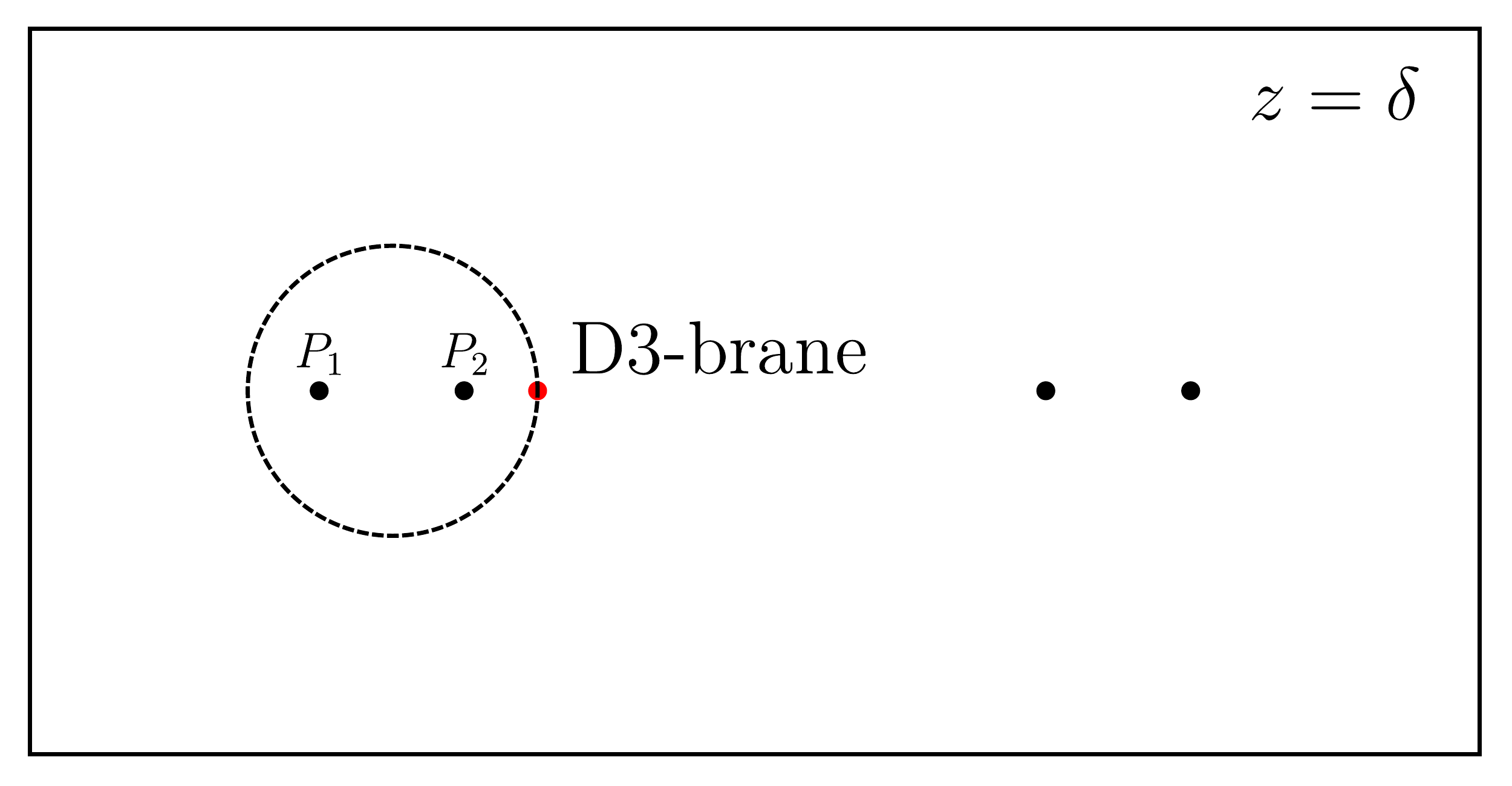}
    \caption{The red dot represents the D3-brane probe. There is a 3-7 string connecting the D3-brane and the non-abelian seven brane at $z = 0$. The dashed circle represents the loop around one pair of the splitting roots $P_1$ and $P_2$ where the $I_1$ locus intersect the $z = \delta$ plane. Note that the precise location of the D3-brane and the shape of the loop are irrelevant as long as they all lie within a small neighborhood of $P_1$ and $P_2$.}
  \label{fig:3-7_string_schematic}
\end{figure*}

While there are multiple ways of interpreting the physics of these extra monodromies, a convenient viewpoint is that of a D3 probe. We consider $3$-$7$ strings stretched from the D3 brane to the 7-brane, and vice versa. The action of the monodromy on these states has a natural interpretation in terms of string junctions, and these states thus serve as building blocks for more general configurations, such as string junctions with ends on multiple 7-branes, which can be obtained by gluing together 3-7 junctions. For a D3 brane at $z = \delta$, with $|\delta|$ small, the 3-7 string from the D3-brane to the non-abelian 7-brane is massive but light, and massless states arise from taking $\delta \rightarrow 0$.The gauge symmetry of the seven-brane is seen as a flavor symmetry from the D3 probe viewpoint, which will allow us to compute the corresponding representation using string junctions on the singular space.  By bringing the D3-brane around the new loops that appears when $\epsilon \neq 0$ one finds an action on the 3-7 strings that does not exist in the undeformed geometry, and the monodromy can be read off using the analytic technique of Section~\ref{sec:general}. We will show that this can induce a monodromy not only on the electromagnetic charge of the 3-7 string under the $U(1)$ carried by the D3-brane, but also the non-abelian flavor representation. In such a case, since the loop can be taken arbitrarily small, the states should be identified and one must quotient by the monodromy action, giving rise to a non-simply-laced flavor algebra on the D3-brane, or alternatively a non-simply-laced gauge algebra on the seven-brane. For type $IV$ and $IV^*$ the picture is schematically shown as in Figure~\ref{fig:3-7_string_schematic}. In Table~\ref{table:s/ns_lie_algebras} we present the cases that will be discussed in this work, which provides more than enough flavor to see the general picture.

\subsection{A simple technique for determining vanishing cycles of $I_1$ fibers}\label{sec:general}

As we saw in the previous section moving from split to non-split fiber above the non-abelian 7-brane along $z = 0$ splits the $I_1$ loci on $z = 0$ into multiple components. As we wish to compute the monodromy action upon encircling the components of this split $I_1$ locus we will derive a simple method to read off the corresponding vanishing cycles. We begin with a general observation on the structure of the roots of the cubic $v(x) := x^3 + fx + g$ that appears in the Weierstrass equation. The roots take the form
\begin{align}
  x_1 &= -\frac{2^\frac{1}{3} f}{3^\frac
    {1}{3}(\sqrt{3} \sqrt{\Delta}-9 g)^\frac{1}{3}} + \frac{(\sqrt{3} \sqrt{\Delta}-9 g)^\frac{1}{3}}{2^\frac{1}{3} 3^\frac{2}{3}}, \label{eq:root1} \\
    x_2 &= \frac{\left(1+i \sqrt{3}\right) f}{2^\frac{2}{3} 3^\frac{1}{3} (\sqrt{3} \sqrt{\Delta}-9 g)^\frac{1}{3}} - \frac{\left(1-i \sqrt{3}\right)
      (\sqrt{3} \sqrt{\Delta}-9 g)^\frac{1}{3}}{2\times2^\frac{1}{3} 3^\frac{2}{3}}, \label{eq:root2} \\
    x_3 &= \frac{\left(1-i \sqrt{3}\right) f}{2^\frac{2}{3} 3^\frac{1}{3}(\sqrt{3}\sqrt{\Delta}-9 g)^\frac{1}{3}} - \frac{\left(1+i \sqrt{3}\right) (\sqrt{3} \sqrt{\Delta}-9 g)^\frac{1}{3}}{2\times2^\frac{1}{3} 3^\frac{2}{3}} \label{eq:root3}.
\end{align}
We immediately see that the discriminant of the elliptic curve $\Delta$ naturally enters the expressions of the roots, which will allow us to simply expressions below.

Let us restrict to a small disc $D$ in the base with complex coordinate $u$, then
$\Delta\in \Gamma(-12K_B)$ becomes a polynomial function  $\Delta_D:= \Delta|_D$ of $u$ that depends on the choice of disc. Then $\Delta_D = (u-u_1)(u-u_2)\cdots(u-u_n)$ where $u_i$'s are the points where the fiber degenerate, and we do not assume that the $u_i$'s are distinct. We will see that  upon looping around any $u_i$, with an $I_1$ fiber above, some pair of roots $\{ x_i, x_j\}$ swaps. We now select an arbitrary $u_i$ with $I_1$ fiber above, say $u_1$, and investigate the behavior of the cycles of the elliptic curve upon carrying it around $u_1$. We consider a change of variables $u_B = u-u_1$ which we parameterize in polar coordinates as $u_B = \lambda e^{i\theta}$. Then $\Delta\sim C_0\lambda e^{i\theta}$ where $\theta \in [0, 2\pi )$ parameterizes a loop around $u_1$ and $C_0$ is a finite term which does not become small. For our purpose it suffices to treat $C_0$ as a constant since any non-constant part of $C_0$ will be of order at least $\sim O(\lambda)$, and we will see our results depends only on the behavior at order $\sim O(\lambda^\frac{1}{2})$ or lower. We will set $C_1 = C_0^\frac{1}{2}$ and $\rho = \lambda^\frac{1}{2}$ for notational convenience, but we will continue to use $\theta$ because it parameterizes the actual loop in the base.

After the substitution and approximation the three roots become:
\begin{equation}\label{eq:xsols_gen}
  \begin{aligned}
    x_1 &= \frac{(-9 g+\sqrt{3} C_1 e^{\frac{i \theta }{2}} \rho )^{1/3}}{\sqrt[3]{2}\times3^{2/3}}-\frac{(\frac{2}{3})^{1/3} f}{(-9g+\sqrt{3} C_1 e^{\frac{i \theta }{2}} \rho )^{1/3}}, \\
      x_2 &= \frac{\left(1+i \sqrt{3}\right) f}{2^{2/3}3^{1/3}(-9 g+\sqrt{3} C_1e^{\frac{i \theta }{2}} \rho )^{1/3}}-\frac{\left(1-i \sqrt{3}\right) (-9 g+\sqrt{3} C_1 e^{\frac{i \theta }{2}} \rho )^{1/3}}{2^{4/3}\times3^{2/3}}, \\
      x_3 &= \frac{\left(1-i \sqrt{3}\right) f}{2^{2/3} 3^{1/3} (-9 g+\sqrt{3} C_1 e^{\frac{i \theta }{2}} \rho)^{1/3}}-\frac{\left(1+i \sqrt{3}\right) (-9 g+\sqrt{3} C_1 e^{\frac{i \theta}{2}} \rho )^{1/3}}{2^{4/3}\times3^{2/3}}.
  \end{aligned}
\end{equation}
Now we can expand with respect to $\rho$\footnote{In general $\epsilon$ is dimensionful, and so we are really expanding in $\epsilon/\epsilon_a$, where the $\epsilon_a$ are the other relevant scales in the Weierstrass model, but we find the same results as simply naively expanding in the parameter $\epsilon$.} and only keep the terms up to $O(\rho)$ so that the above roots are further simplified to:
\begin{align*}
  x_1 =& -\frac{2^{1/3} f}{3 (-g)^{1/3}}+\left(\frac{-g}{2}\right)^{1/3}+\frac{C_1 e^{\frac{i \theta }{2}} \left(2 f (-g)^{1/3}-3 (2)^{1/3} g\right)}{27\times2^{2/3} \sqrt{3}(-g)^{5/3}}\rho+O(\rho^2), \\
    x_2 =& \frac{3 i 2^{2/3}
      \left(\sqrt{3}+i\right) (-g)^{1/3} g-2 i (2)^{1/3} \left(\sqrt{3}-i\right) f (-g)^{2/3}}{12 g} \\
      &+\frac{C_1 e^{\frac{i \theta }{2}} \left(3(2)^{1/3} \left(1-i \sqrt{3}\right) g-2 i \left(\sqrt{3}-i\right) f (-g)^{1/3}\right)}{54\times2^{2/3} \sqrt{3} (-g)^{5/3}}\rho+O(\rho^2), \\
    x_3 =& \frac{2 i (2)^{1/3} \left(\sqrt{3}+i\right) f (-g)^{2/3}+3 i 2^{2/3} \left(\sqrt{3}-i\right) (-g)^{4/3}}{12 g} \\
      &+ \frac{C_1 e^{\frac{i \theta }{2}} \left(2 i \left(\sqrt{3}+i\right) f(-g)^{1/3}+3 (2)^{1/3} \left(1+i \sqrt{3}\right) g\right)}{54\times2^{2/3} \sqrt{3}(-g)^{5/3}}\rho+O(\rho^2).
\end{align*}
In fact, it will be sufficient to concentrate on
the $O(1)$ parts of these three expressions, which
we refer to as $A_1$, $A_2$ and $A_3$, respectively.

As an example consider $A_1 - A_2$, which sets the $O(1)$ distance between $x_1$ and $x_2$. We have:
\begin{equation}\label{eq:DS12}
  A_1 - A_2 = \frac{2^{4/3} \left(-3- i \sqrt{3}\right) f+3\times2^{2/3} \left(3-i \sqrt{3}\right) (-g)^{2/3}}{12 (-g)^{1/3}}\, .
\end{equation}
Recall that $\Delta_D = (4f^3 + 27g^2)|_D$ and we are expanding around one of the zero points of $\Delta_D$, and so $\Delta_D \sim O(\lambda)=O(\rho^2)$. Since this is smaller than the $\rho$-dependent parts of the roots in Equation~\ref{eq:xsols_gen} we can set $4f^3 = -27g^2$ to leading order, without losing the leading $\rho$-dependence of the roots. Hence we can solve the equation $4f^3 = -27g^2$ for $f$, and the solutions take the form of Eq. \ref{eq:froot1} - \ref{eq:froot3}. For instance, one can explicitly check that, when Equation~(\ref{eq:froot3}) holds, we have $A_1 - A_2 = 0$, and hence we obtain $x_1 - x_2 \sim \rho e^{\frac{i\theta}{2}}$.

There are two crucial pieces of information that we wish to extract from the above result.
The first is to notice that the limit $\rho\to 0$ corresponds to approaching the $I_1$ seven-brane at $z = u_1$, where $x_1\rightarrow x_2$; that is, $x_1$ and $x_2$ collide.
The second is encoded in the $e^{\frac{i\theta}{2}}$ part. While the small parameter $\rho$ sets the order such that the above approximations can be systematically performed, the taking $\theta$ from zero to $2\pi$ swaps $x_1$ and $x_2$; such a swap occurs when the elliptic curve is brought around a small circle centered at the point satisfying Equation (\ref{eq:froot3}) . This swap corresponds precisely to the Picard-Lefschetz monodromy of traversing the loop around this $I_1$  locus, and this technique gives an efficient way to read off the vanishing cycle and compute the corresponding monodromy matrix. A similar analysis may be performed for the $x_1$-$x_2$ swap and the $x_2$-$x_3$ swap. The former happens when the elliptic curve is brought around the point where Eq.(\ref{eq:froot1}) holds while the latter at where Eq.(\ref{eq:froot2}) holds.
\begin{table*}[t!]
\centering
\begin{tabular}{ c || c | c | c | c | c | c }
Fiber type & $I_4$ & $IV$ & $IV^*$ & \multicolumn{2}{c|}{$I_0^*$} & $I_1^*$ \\
\hline
Split & $SU(4)$ & $SU(3)$ & $E_6$ & \multicolumn{2}{c|}{$SO(8)$} & $SO(10)$ \\
\hline
Outer-automorphism & $\mathbb{Z}_2$ & $\mathbb{Z}_2$ & $\mathbb{Z}_2$ & $\mathbb{Z}_2$ & $\mathbb{S}_3$ & $\mathbb{Z}_2$ \\
\hline
Non-split & $Sp(2)$ & $Sp(1)$ & $F_4$ & $SO(7)$ & $G_2$ & $SO(9)$
\end{tabular}
\caption{Simply-laced and non-simply-laced Lie algebras that are analyzed in this paper.}
\label{table:s/ns_lie_algebras}
\end{table*}

\subsection{Monodromy action in examples}\label{sec:cases_analytic}

In this section we will study concrete examples where the fiber of a non-abelian 7-brane becomes non-split via a deformation. Our analysis in this section will be the relevant geometric analyses for the reduction of simply-laced gauge algebras to  non-simply laced ones; this fact, and our naming conventions for each example, will be justified in the corresponding subsections in Section~\ref{sec:junc_analysis}. We begin with the most computationally tedious example, which will allow us to demonstrate our technique in full detail.

\subsubsection{Type $I_{0s}^*$: $SO(8)\rightarrow G_2$}\label{sec:G2_analytic}

To obtain the Weierstrass model for $G_2$ we will start from the Weierstrass model for $SO(8)$. The latter is obtained when $x^3 + fx +g$ factorizes into three pieces. We can in general let 
\begin{equation}\label{eq:SO8Wmodel}
	x^3 + fx +g = (x+Bz+Cz)(x-Bz)(x-Cz).
\end{equation}
So that:
\begin{equation}
  \begin{aligned}
	f &= -B^2 z^2 -BCz^2 -C^2z^2 \\
	g &= B^2 C z^3+B C^2 z^3
  \end{aligned}
\end{equation}
Here and henceforth capital letters denote generic holomorphic functions of $z$ and $t$ in the local geometry, whose precise forms are not crucial, and are example-dependent. To break $SO(8)$ to $G_2$ model we can simply add a term to both $f$ and $g$ so to so that the LHS of Eq.~\ref{eq:SO8Wmodel} can no longer be factorized. The simplest such terms for $f$ and $g$ are the terms that are linear in $\epsilon$ and vanish to order 2 and 3 in $z$, respectively. Hence the Weierstrass model for $SO(8)\to G_2$ is:\begin{equation}\label{model:G2}
  \begin{aligned}
    f &= -B^2 z^2-B C z^2-C^2 z^2+F_1 z^2 \epsilon +F_2 z^3, \\
    g &= B^2 C z^3+B C^2 z^3+G_1 z^3 \epsilon +G_2 z^4,
  \end{aligned}
\end{equation}
where the gauge group is $SO(8) $for $\epsilon=0$ and $G_2$ for  $\epsilon\neq 0$. Taking $z = \delta \neq 0$, the residual discriminant $\Delta_R$ takes the form:
\begin{equation}
  \Delta_R(t; \delta; \epsilon) = 27 \left(B^2 C+B C^2+G_1 \epsilon +\delta G_2\right)^2-4 \left(B^2+B
      C+C^2-F_1 \epsilon -\delta F_2\right)^3
\end{equation}
Solving $\Delta_R(t; 0; 0) = 0$ for $B$, there are three double roots at $B = -2C$, $B = -\frac{1}{2}C$, $B = C$. After turning on $\epsilon$, a direct computation shows that each of the three double roots splits to a pair of simple roots.\footnote{We will often avoid listing roots such as these because of the lengthy nature of the expressions.} Since all of the roots split, one expects that the geometric monodromy action could be different for each pair, which may be important in obtaining a larger group of automorphisms of $SO(8)$. In this case turning on $\delta$ does not introduce further splitting due to the fact that after turning on $\epsilon$ each of the three double roots already splits into two simple roots, and hence no further splitting can occur, due to the structure of the polynomial $\Delta_R$.

To compute the monodromy, we will apply the method derived in section \ref{sec:general}. There are three pairs of solutions to the equation $\Delta_R(t; \delta; \epsilon) = 0$. If we approach an arbitrary root $B_R$ of $\Delta_R(t; \delta; \epsilon) = 0$, the solutions to the equation $x^3 + fx + g = 0$ become:
\begin{equation}\label{eq:xsols_G2}
  \begin{aligned}
    x_1 &= \frac{\sqrt[3]{\sqrt{B_t} W+V}}{3\times\sqrt[3]{2}}-\frac{\sqrt[3]{2} U}{3 \sqrt[3]{\sqrt{B_t} W+V}}, \\
    x_2 &= \frac{\left(1+i \sqrt{3}\right)U}{3\ 2^{2/3} \sqrt[3]{\sqrt{B_t} W+V}}-\frac{\left(1-i \sqrt{3}\right) \sqrt[3]{\sqrt{B_t} W+V}}{6 \sqrt[3]{2}}, \\
      x_3 &= \frac{\left(1-i\sqrt{3}\right) U}{3\times2^{2/3} \sqrt[3]{\sqrt{B_t} W+V}}-\frac{\left(1+i \sqrt{3}\right)\sqrt[3]{\sqrt{B_t}W+V}}{6\sqrt[3]{2}} 
  \end{aligned}
\end{equation}
where $U = -3 B^2 \delta^2-3 B C \delta^2-3 C^2 \delta^2+3 \delta^2 F_1 \epsilon +3 \delta^3 F_2$, $V = -27 B^2 C \delta^3-27 B C^2 \delta^3-27 \delta^3 G_1 \epsilon - 27 \delta^4 G_2$, $B_t \equiv B - B_R = \rho e^{i\theta}$, $\rho$ is a small parameter, and $W \propto \delta^3$ is a complicated function whose exact form is irrelevant. 

Before we perform any perturbative expansions, it is easy to see that $x_1$, $x_2$ and $x_3$ are all proportional to $\delta$. Since $z \sim\delta\equiv\delta_0 e^{i\theta}$, we know immediately that bringing the D3-brane along a loop around the $z = 0$ plane, i.e., the stack of the 7-branes, will lead to a geometric monodromy action that corresponds to a $2\pi$ rotation of the three roots. This $2\pi$ rotation corresponds to a monodromy $-I_{2\times 2}$ matrix acting on $(p,q)$-cycles, or alternatively the $(p,q)$-charges of string junctions. The reason for this is that the orientation of the 1-cycles is determined in the double cover of the $x$-plane, and so a $2\pi$ rotation in the $x$-plane corresponds to a $\pi$ rotation in the double cover, which reverses the orientation of the 1-cycles in the double cover. This fact which will play an important role in our subsequent discussions; see \cite{Grassi:2016bhs} for a detailed discussion.

We now expand the above solution with respect to $\rho$. Keeping only the lowest order in $\rho$ we obtain:
\begin{equation}\label{eq:sols_x_expanded}
  \begin{aligned}
    \frac{x_1}{\delta} &= \frac{2^{2/3} V^{2/3}-2 \sqrt[3]{2} U}{6\sqrt[3]{V}} + \sqrt{\rho}e^{\frac{i}{2}\theta} \frac{W \left(2 \sqrt[3]{2} U+2^{2/3} V^{2/3}\right)}{18 V^{4/3}}, \\
      \frac{x_2}{\delta} &= \frac{\left(1+i \sqrt{3}\right) U}{3\times2^{2/3}\sqrt[3]{V}}-\frac{\left(1-i \sqrt{3}\right) \sqrt[3]{V}}{6\sqrt[3]{2}} + \sqrt{\rho}e^{\frac{i}{2}\theta} \left(-\frac{\left(1+i\sqrt{3}\right) U W}{9\times2^{2/3} V^{4/3}}-\frac{\left(1-i\sqrt{3}\right) W}{18 \sqrt[3]{2}V^{2/3}}\right), \\
      \frac{x_3}{\delta} &= \frac{\left(1-i\sqrt{3}\right) U}{3\times2^{2/3} \sqrt[3]{V}}-\frac{\left(1+i\sqrt{3}\right) \sqrt[3]{V}}{6 \sqrt[3]{2}} + \sqrt{\rho}e^{\frac{i}{2}\theta} \left(-\frac{\left(1-i \sqrt{3}\right) U W}{9\times2^{2/3} V^{4/3}}-\frac{\left(1+i \sqrt{3}\right) W}{18 \sqrt[3]{2}V^{2/3}}\right).
  \end{aligned}
\end{equation}

It is easy to show that $4U^3 + V^2 \propto \Delta_R(B_t; \delta; \epsilon) \sim O(\rho)$, and so at leading order in $O(\rho^\frac{1}{2})$, $4U^3 + V^2 = 0$. There are three solutions to this equation:
\begin{align}
  U_1 &= -\left(-\frac{1}{2}\right)^{2/3} V^{2/3}, \label{eq:G2U1} \\
  U_2 &= -\frac{V^{2/3}}{2^{2/3}}, \label{eq:G2U2} \\
  U_3 &= \frac{\sqrt[3]{-1} V^{2/3}}{2^{2/3}}. \label{eq:G2U3}
\end{align}
We now substitute these relations between $U$ and $V$ back into Eq.(\ref{eq:sols_x_expanded}) and investigate the behavior of the set of roots $\{x_1, x_2, x_3\}$. If $B_R\in B$ is a solution to $\Delta_R(t; \delta; \epsilon) = 0$ where the relation given by Eq.(\ref{eq:G2U1}) holds, then $x_1 - x_2 \propto \sqrt{\rho}e^{\frac{i}{2}\theta}$, and so dragging the D3-brane probe along a loop around such a $B_R$ induces a swap of the roots $x_1$ and $x_2$. By the same logic, it is easy to show that when $B_R$ is a solution such that Eq.(\ref{eq:G2U2}) holds there is an $x_2$-$x_3$ swap and when $B_R$ is such that Eq.(\ref{eq:G2U3}) holds there is an $x_1$-$x_3$ swap. Recall that in each case there is also an overall
$2\pi$ rotation, in addition to the swap.

\begin{figure*}[ht]
  \centering
  \includegraphics[width=.5\linewidth]{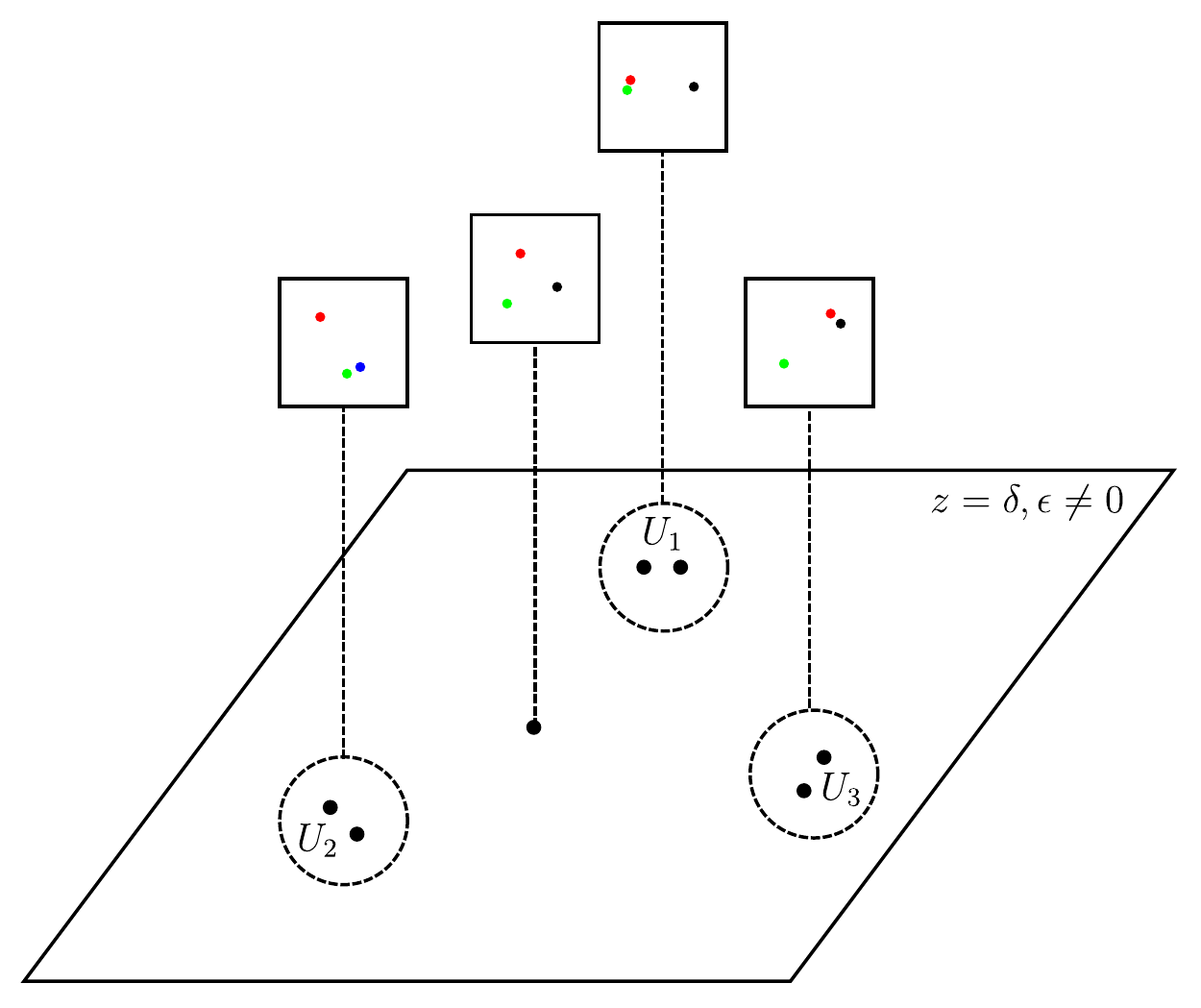}
    \caption{A picture of the $G_2$ case for the elliptic fibration over the $z = \delta$ plane while $\epsilon\neq 0$. $U_1$, $U_2$ and $U_3$ are three regions such that Eq.(\ref{eq:G2U1}), (\ref{eq:G2U2}) and (\ref{eq:G2U3}) hold to the order $O(\rho^\frac{1}{2})$, respectively. In these regions the fiber becomes nearly degenerate. The red dot represents the root $x_1$, the green dot $x_2$ and the blue dot $x_3$. The central node on the $z = \delta$ plane is at arbitrary position as long as it is away from the region where the fiber becomes degenerate.}
  \label{fig:G2_illustration}
\end{figure*}

One possible complication is whether these are indeed three different swaps since, e.g., an $x_1$-$x_2$ swap in some local patch may become, say, an $x_1$-$x_3$ swap in some other local patch if there there is a nontrivial transition function between patches. However, our entire analysis is a local one, and the (arbitrary) chosen ordering of the three roots $\{x_1, x_2, x_3\}$ does not change when moving around on the $z = \delta$ plane. It is simple to verify this numerically. The configuration is schematically shown in Figure~\ref{fig:G2_illustration}.

Note that in this case all the three possible swaps between $\{x_1, x_2, x_3\}$ are realized, and therefore the precise correspondence between particular loops and vanishing cycles is not important for computing monodromy actions on string junctions. This will not be the case in general, and we will see that the explicit correspondence between swaps, and the relations between $U$ and $V$, are crucial in the string junction computations.

\subsubsection{Type $I_{0ss}^*$: $SO(8)\rightarrow SO(7)$}\label{sec:SO7analytic}

After analyzing the $SO(8)\rightarrow G_2$ case it is natural to consider monodromy reduction $SO(8)\rightarrow SO(7)$.

The Weierstrass model takes the form:
\begin{align}
  f & = A z^2 \epsilon -B^2 z^2-B C z^2-C^2 z^2+F z^3,  \\
  g &= A B z^3 \epsilon +A C z^3 \epsilon +B^2 C z^3+B C^2 z^3,
\end{align}
for which:
\begin{equation}
  \Delta_R(t; \delta; \epsilon) = 27 (B+C)^2 (A \epsilon +B C)^2-4 \left(-A \epsilon +B^2+B C+C^2-\delta  F\right)^3.
\end{equation}

The roots of $x^3 + fx + g = 0$ can be written in the same form as in the $G_2$ case, given in Equation~\ref{eq:xsols_G2}. In this case  $U$, $V$ and $W$ take the form:
\begin{align*}
  U &= 3 A \delta ^2 \epsilon -3 B^2 \delta ^2-3 B C \delta ^2-3 C^2 \delta ^2+3 \delta ^3 F, \\
  V &= -27 A B \delta ^3 \epsilon -27 A C \delta ^3 \epsilon -27 B^2 C \delta ^3-27 B C^2 \delta ^3.
\end{align*}
Here $W\sim\delta^3 W_0$, where the precise form of $W_0$ is unimportant. Again there an overall multiplicative factor of $\delta$  in the $x_i$, which indicates the presence of an $-I_{2\times 2}$ monodromy action.

The difference between the $SO(7)$ case and $G_2$ case lies in the pattern of how the roots of $\Delta_R(t; \delta; \epsilon) = 0$ split after turning on $\epsilon$. Let us focus on the $z = 0$ plane. There are again three double roots of $\Delta_R(t; 0; 0) = 0$ at $B = -2C$, $B = -\frac{1}{2}C$ and $B = C$. After turning on $\epsilon$ only the double root at $B = C$ split into two distinct roots $B = C \pm \sqrt{A\epsilon}$. The other two double roots move a small amount but remain degenerate, i.e. they do not split. Therefore, the still-degenerate double roots do not contribute additional monodromy action on the string junctions.

Again we can consistently set $4U^3 + V^2 = 0$ to leading order. The three solutions to this equation again correspond to a swap of $x_1$-$x_2$, $x_2$-$x_3$ or $x_1$-$x_3$, respectively; however, only one of the three swaps is relevant because the pair of the roots that provide a new geometric monodromy action are at $B = C \pm \sqrt{A\epsilon}$. Substituting these expressions of $B$ into the expressions of $U$ and $V$, we find that the corresponding relation is given by Equation~(\ref{eq:G2U3}), which is realized at both of these two new roots. Therefore, the geometric monodromy action induced by looping around one of these two roots corresponds to an $x_1$-$x_3$ swap. Of course, there is a freedom of relabeling $x_1$, $x_2$ and $x_3$, so that when we say the geometric monodromy action corresponds to an $x_1$-$x_3$ swap, we have fixed the labeling by a choice of vanishing cycle along the non-abelian 7-brane, as will be made clear in our discussion of string junctions in Section~\ref{sec:junc_analysis}.

\subsubsection{Type $IV$: $SU(3)\rightarrow Sp(1)$}\label{sec:IVanalytic}

Let us now consider consider type $IV_s$ and $IV_{ns}$ fibers. The Weierstrass model takes the form:
\begin{align*}
  f &= F z^2, \\
  g &= z^2 \left(G_1 \epsilon +G_2 z+M^2\right) 
\end{align*}
for which we have:
\begin{equation}
  \Delta_R(t; \delta; \epsilon) = 4\delta^2 F^3+27 \left(G_1\epsilon +\delta G_2 + M^2\right)^2
\end{equation}
Solving for the roots of $x^3 + fx +g = 0$ we again find the same structure, $U$, $V$ and $W$ defined as:
\begin{equation}\label{eq:IVUV}
  \begin{aligned}
    U &= 3F\delta^2, \\
    V &= 27(-\delta^2 G_1 \epsilon - G_2\delta^3 -\delta^2 M^2).
  \end{aligned}
\end{equation}
Again we have $W\sim\delta^3 W_0$. Immediately we can see a difference between this case and the $G_2$ and $SO(7)$ cases: there is no overall multiplicative $\delta$ factor for each root. Instead, the leading order overall multiplicative factor is $\delta^\frac{2}{3}$. This represents an overall rotation of $\{x_1, x_2, x_3\}$ upon traversing a loop around the $z = 0$ plane; instead of a $2\pi$ rotation, we find a $\frac{4}{3}\pi$ rotation. There is still the relation $4U^3 + V^2 = 0$ and each of its three solutions corresponds to one of the three different swaps. However, unlike the previous cases, there is an ambiguity in which of the three swaps is realized at different components of $\Delta_R(t; \delta; \epsilon) = 0$.

In this case, there is a fourfold root $M = 0$ to the equation $\Delta_R(t; 0; 0) = 0$. After turning on $\epsilon$, the fourfold root splits into to two double roots $M = \pm i \sqrt{G_1\epsilon}$. After turning on $\delta$ each of the double roots splits further into two simple roots. These four simple roots are located at 
\begin{equation}
	M = \pm (\pm\frac{2iF^\frac{2}{3}\delta }{3\sqrt{3}}- G_1\epsilon - G_2\delta)^\frac{1}{2}.
\end{equation}
Substituting these expressions of $M$ into Equation~\ref{eq:IVUV}, it is straightforward to see which of relations between $U$ and $V$ are realized. The relations realized at $M = \pm (\pm\frac{2iF^\frac{2}{3}\delta }{3\sqrt{3}}- G_1\epsilon - G_2\delta)^\frac{1}{2}$ are $U_3$, $U_1$, $U_3$ and $U_1$, for the combinations $\{+,+ \}$, $\{+,- \}$, $\{-,+ \}$, and $\{-,- \}$, respectively, where $U_1 = -\left(-\frac{1}{2}\right)^\frac{2}{3} V^\frac{2}{3}$ and $U_3 = \frac{\sqrt[3]{-1}}{2^\frac{2}{3}}V^\frac{2}{3}$. One can show that $U_3$ corresponds to an $x_1$-$x_3$ swap and $U_1$ to $x_1$-$x_2$ swap, so that for each pair of the simple roots, which in the $\delta = 0$ limit recombines back into a double root, the monodromy action is a combination of an $x_1$-$x_3$ swap and an $x_1$-$x_2$ swap.

In a string junction analysis, if the $SL(2, \mathbb{Z})$ matrix acting on the asymptotic charges of the string junctions corresponding to $x_1$-$x_2$ swap is $M_1$ and to $x_1$-$x_3$ is $M_3$, then the total monodromy matrix acting on the asymptotic charge is $M_1\cdot M_3$. We will make this point clearer in Section~\ref{sec:junc_analysis}.

\subsubsection{Type $IV^*$: $E_6\rightarrow F_4$}\label{sec:IV*analytic}

This case is structurally very similar to the type $IV$ case in the above section. The Weierstrass model is:
\begin{align*}
  f &= Fz^3, \\
  g &= z^4(\epsilon G_1 + \delta G_2 +M^2)\, ,
\end{align*}
for which we have:
\begin{equation}
  \Delta_R(t; \delta; \epsilon) = 4\delta F^3 + 27 G_1^2 \epsilon^2 + 54\delta \epsilon G_1 G_2 + 54 G_1 M^2 \epsilon +27\delta^2 G_2^2+ 54\delta G_2 M^2 + 27 M^4.
\end{equation}
It is straight forward to show that pattern of the splitting of the roots is the same as in the type $IV$ case, and the geometric monodromy actions realized in this case are also identical to those of the type $IV$ case. The only difference that lies between type $IV^*$ and $IV$ is that in this case there is an overall multiplicative $\delta^\frac{4}{3}$ factor instead of a $\delta^\frac{2}{3}$ factor, and so upon encircling the $z = 0$ plane there is an overall $\frac{8}{3}\pi$ rotation. The rest if the analysis is identical to the $IV_s \rightarrow IV_{ns}$ case.

\subsubsection{Type $I_1^*$: $SO(10)\rightarrow SO(9)$}\label{sec:I1*_junc_analytic}

In this case we use a Tate model to obtain the Weierstrass model:
\begin{align*}
  f =& \  -\frac{1}{48} A_1^4 z^4+\frac{1}{48} z^3 \left(-8 A_1^2 A_2+24 A_1 A_3+48 A_4\right)-\frac{A_2^2 z^2}{3}, \\
  g =& \  \frac{A_1^6 z^6}{864}+\frac{1}{864} z^4 \left(48 A_1^2 A_2^2-144 A_1 A_2 A_3-288 A_2 A_4+216 A_3^2+864 A_{64} \epsilon \right) \\
    &+\frac{1}{864} z^5 \left(12 A_1^4 \text{A2}-36 A_1^3 A_3-72 A_1^2 A_4+864 A_{65}\right)+\frac{2 A_2^3 z^3}{27}.
\end{align*}
As usual the roots of $x^3 + fx + g = 0$ have the same structure as before, although with a different form of $U$, $V$ and $W$.
After turning on $\epsilon$ the roots of $\Delta_R(t; \delta; \epsilon) = 0$ split from a double root at $A_3 = 0$ to two simple roots at $A_3 = \pm 2i \sqrt{A_{64}\epsilon}$. However, there is a subtlety in this model: after turning on $\delta$, there are two more roots appearing because of the presence of the higher order terms in $A_3$ when $\delta \neq 0$. These two extra roots does not introduce any additional monodromy, as they do not exist in the limit $\delta \rightarrow 0$.

It is easy to show that there is an overall multiplicative $\delta$ factor in the roots of $x^3 + fx + g = 0$, which corresponds to a $2\pi$ rotation, as before. It also not hard to show that $x_1 - x_3 \propto \sqrt{\delta}$ so that there is also an extra $x_1$-$x_3$ swap while looping around the $z = 0$ plane. Proceeding as before, it is not hard to show it is $U_3$ that is realized in the neighborhood of one of the relevant simple roots of $\Delta_R(t; \delta; \epsilon) = 0$, and so that in this case the geometric monodromy action is an $x_1$-$x_3$ swap.

\subsubsection{Type $I_4$: $SU(4)\rightarrow Sp(2)$}\label{sec:I4_geom_analytic}

Finally we consider type $I_4$. The Weierstrass model is:
\begin{align*}
  f &= -6 P z\left(G \epsilon +M^2\right)-3 \left(G \epsilon+M^2\right)^2-3 P^2 z^2, \\
  g &= 6 P^2 z^2 \left(G \epsilon +M^2\right)+6 P z \left(G \epsilon +M^2\right)^2+2 \left(G \epsilon +M^2\right)^3+2 P^3 z^3+Q z^4,
\end{align*}
for which we have:
\begin{align*}
  \Delta_R(t; \delta; \epsilon) =& \  4 G^3 \epsilon^3+12 \delta G^2 P \epsilon^2+12 M^4(G \epsilon +\delta P)+12 M^2 (G\epsilon +\delta P)^2 \\
  &+12 \delta^2 G P^2 \epsilon +4 M^6+4\delta^3 P^3+\delta^4 Q.
\end{align*}
The solutions to $x^3 + fx + g = 0$ have the same structure as before. In this case, we have:
\begin{align*}
  U =& \  -G^2 \epsilon^2-2 G M^2 \epsilon -2 \delta G P \epsilon -M^4-2 \delta  M^2 P-\delta^2 P^2, \\
  V =& \  2 G^3 \epsilon^3+6 G^2 M^2 \epsilon^2+6 \delta  G^2 P \epsilon^2+6 G M^4 \epsilon +12 \delta G M^2 P \epsilon \\
  &+6 \delta ^2 G P^2 \epsilon +2M^6+6 \delta M^4 P+6 \delta^2 M^2 P^2+2 \delta^3 P^3+\delta ^4 Q,
\end{align*}
and again $4U^3 + V^2$ vanishes to leading order.

In this case it can easily be shown that there is no multiplicative $\delta$ factor in the solutions to $x^3 + fx + g = 0$, and so there is no overall rotation of the configuration of the roots when bringing the elliptic fiber along a loop around the $z = 0$ plane. A higher order analysis shows that $x_1 - x_3 \propto \delta^2$, and therefore although there is no overall rotation of $\{x_1, x_2, x_3\}$, $x_1$ rotates around $x_3$ by $4\pi$, and so $x_1$ and $x_3$ get swapped four times in this process.

In this case the solution to $\Delta_R(t; 0; 0) = 0$ is a sixfold root $M = 0$. Turning on $\epsilon$, $M = 0$ splits into two threefold roots $M = \pm i \sqrt{G\epsilon}$. Upon turning on $\delta$ each of the threefold roots splits further into three distinct roots. In the order we chose, the $U$ roots that are realized are $U_1$, $U_3$ and $U_2$. Therefore the swaps on the $x$ roots are first an $x_1$-$x_2$ swap, then an $x_1$-$x_3$ swap, and finally an $x_2$-$x_3$ swap. The combined monodromy action is then an $x_1$-$x_3$ swap, but this is not the whole story. Recall the swap is realized not via a $\mathbb{Z}_2$ action, but instead via a $\pi$ rotation of the two roots involved. One can then see that there is also an overall $2\pi$ rotation in addition to the $x_1$-$x_3$ swap upon encircling a loop enclosing the three roots that, in the $\delta = 0$ limit ,recombine into a single threefold root. 

This presence of this overall $2\pi$ rotation is manifest in the $\delta = 0$ limit while keep $\epsilon$ non-zero. In this limit the $x$ roots consists of a simple root $x_s = -2(M^2 + G\epsilon)$ and a double root $x_d = M^2 + G\epsilon$. In this limit the solutions to $\Delta_R(t; \delta; 0) = 0$ are $M = \pm i\sqrt{G\epsilon}$, and so that by the same logic as in the $\delta \neq 0$ case we can still expand the $x$ roots around one of these two solutions. We therefore let $M = \rho e^{i\theta} \pm i\sqrt{G\epsilon}$ where $\rho$ is a small parameter, and keep only the lowest order terms in $\rho$. We obtain $x_{1, 3} = 2i \rho e^{i\theta}\sqrt{G\epsilon}$, $x_2 = -4i \rho e^{i\theta}\sqrt{G\epsilon}$, of which the relevant factor is $\rho e^{i\theta}$. This demonstrates there is an overall $2\pi$ rotation of the configuration of $x$ roots. Note in this limit, the $x_1$-$x_3$ swap can not readily be seen, but the presence of the overall $2\pi$ rotation is much more transparent. \\ 

\textit{Remark} We want to point out a fact that will be important in the string junction analysis later that, in both the $I_1^*$ and the $I_4$ cases, it is the same $x_1$-$x_3$ swap that appears in both the monodromy action corresponding to a loop around the $z = 0$ plane and around a loop enclosing one of the relevant simple roots of $\Delta_R(t; \delta; \epsilon) = 0$ in the $I_1^*$ case, or one of the two pairs of the splitting threefold roots of $\Delta_R(t; \delta; \epsilon) = 0$ when turning on $\epsilon$ in the $I_4$ case. We will see this piece of information is extremely useful in obtaining the physical consequence of the geometric monodromy action, i.e., the corresponding $SL(2, \mathbb{Z})$ matrix acting on the asymptotic charges of the string junctions.

\section{String Junctions, Monodromy, and Non-simply-laced Algebras}\label{sec:junc_analysis}

Having thoroughly analyzed the geometric monodromy in a number of examples, we now discuss the physical implications. The central qualitative fact derived in the previous section is that
the $\epsilon$-deformations, which transition between Kodaira split fibers
and Kodaira non-split (or semi-split) fibers, can split $I_1$ loci in the vicinity of the
seven-brane, creating new loops that may be traversed.
These new loops allow for new monodromy actions that were not present prior to the $\epsilon$ deformation, and we now analyze that monodromy action on the symmetry algebra.

In particular, we consider a D3-brane that can traverse those loops. Upon returning to its original position the monodromy action may give rise to an action, and thus reduction, of its 3-7 string spectrum. The flavor symmetry  $G$ of the 3-7 strings corresponds to the gauge symmetry on the non-abelian seven-brane, and we will see that the monodromy induces a non-trivial map on the representations realized by 3-7 strings, which in turn induces an outer automorphism on $G$. We will demonstrate this in all of our examples, and in all cases the result matches known results from the M-theory Coulomb branch description of F-theory. We emphasize, however, that we obtain the results on the singular space, without deformation or resolution to a smooth variety. Our methods will be partially justified below, and will be fully justified in \cite{bcjt}.

\subsection{String Junctions on Deformed Spaces}

We begin by reviewing the now standard story of string junctions on deformed elliptic
fibrations. See~\cite{Gaberdiel:1997ud, DeWolfe:1998bi, DeWolfe:1998zf} for early physics
work on string junctions, \cite{Grassi:2013kha, Grassi:2014sda,Grassi:2014zxa} for 
realizations in explicit Weierstrass models, based on a rigorous
geometric and topological treatment \cite{Grassi:2014ffa}.

\vspace{1cm}
\noindent \emph{Geometric setup.}

Consider a Calabi-Yau elliptic fibration $X$ as defined in
Section \ref{sec:geom_monodromy}. Recall that there is a projection map
\begin{equation}
X \xrightarrow{\pi} B
\end{equation}
and that it may be written as a Weierstrass model
\begin{equation}
y^2 = x^3 + f x + g\, ,
\end{equation}
where $f\in \Gamma(\cO(-4K_B))$ and $g\in \Gamma(\cO(-6K_B))$.
The discriminant of the cubic $v_3(x)=x^3+fx + g$ in $x$ is 
\begin{equation}
\Delta = 4f^3 + 27g^2\, .
\end{equation}
We choose a point $p\in B$ such that $E_p:= \pi^{-1}(p)$ is a smooth
elliptic curve. By studying the roots of the cubic, as discussed in
Section \ref{sec:geom_monodromy}, we may easily define a basis on
$H_1(E_p,\mathbb{Z})$.
Neither the section nor the Weierstrass equation are necessary
for the existence of the string junction picture 
\cite{Grassi:2014ffa}, but it
does help facilitate computations.

Suppose $X$ is a small deformation away from a model with non-$I_1$
Kodaira fiber at the locus $\{z=0\}\subset B$, which itself has only
$I_1$ fibers. The $N$ $(p,q)$ seven-branes ($I_1$ fibers) coalesce into the
non-abelian seven-brane when the deformation is turned off. Let us call
the locations of these $I_1$ fibers $p_i$, with $i=1,\dots,N.$
It is then natural to choose $p$ at $z=0$ and to compute
the vanishing cycles of the $I_1$ fibers by following straight line
paths from $p$ to $p_i$. Given an ordering of loops around the $p_i$
that are topologically equivalent to a loop around the whole
configuration determines an ordered set of vanishing cycles
\begin{equation}
Z = \{\gamma_1,\cdots,\gamma_N\}\, .
\end{equation}
Following $\gamma_i$ from the $p_i$
back to $p$ creates a cigar, or Lefschetz thimble or ``prong", in the geometry,
which define elements
\begin{equation}
\Gamma_i \in H_2(X,E_p)\, .
\end{equation} 
We can also take linear combinations of the prongs
\begin{equation}
J = \sum_I J_i \Gamma_i \in H_2(X,E_p)\, ,
\end{equation}
and these objects are string junctions, which can be thought heuristically
as linearly combinations of the prongs. Of course, such objects, which are chains
in the geometry that may have a boundary at $E_p$, can also be defined for
any point $p$, as long as the fiber above it is a smooth elliptic curve.

We must also discuss boundaries and a pairing.
The boundary map is known as an asymptotic charge,
\begin{align}
a: H_2(X,E_p) \to H_1(E_p,\bZ)\, , \qquad a(J)=\partial J\, . 
\end{align} 
The asymptotic charge of each prong is the vanishing cycle
\begin{equation}
\gamma_i := a(\Gamma_i) \in H_1(E_p,\bZ)\, .
\end{equation}
Following successive loops from  $p$ around each of the
$p_i$ determines an ordered set of vanishing cycles.
Now suppose that $X$ is a surface. In this case there is a natural pairing
\begin{equation}
(\cdot,\cdot): H_2(X,E_p) \times H_2(X,E_p) \to \bZ\, ,
\end{equation}
that becomes the topological intersection product on closed classes. Given
the ordered set of vanishing cycles $Z$, this may be computed as described
in~\cite{Grassi:2013kha,Grassi:2014sda,Grassi:2014ffa}. We will simply present the results in examples.

This topological structure is physically relevant.
If a D3-brane is at $p$, then $p$ gains a physical meaning via the worldvolume
theory on the D3-brane. The asymptotic charge of a string
junction ending on the D3-brane is the charge under of the junction
under the $U(1)$ on the D3-brane. These 3-7 strings may also be in representations
of the Lie algebra $G$ associated with the deformed Kodaira fiber, to which we
now turn.

\vspace{1cm}
\noindent \emph{Representation theory.}

Let us recall some basic facts of the representation theory associated with string
junctions. First, for any deformation of the type described, with a non-abelian seven-brane whose Kodaira
fiber has associated simply-laced algebra $\mathfrak{g}$, there is a distinguished set of junctions
\begin{equation}
R := \{J \in H_2(X,E_p) \,\, | \,\, (J,J)=-2, \,\,\, a(J)=0\}.
\end{equation}
that has
\begin{equation}
|R| = dim(\mathfrak{g})-rk(\mathfrak{g}).
\end{equation}
Closer inspection shows that there are natural decomposition into ``positive'' and
``negative'' elements of $R$, and there always exists a set
\begin{equation}
SR=\{\alpha_1,\dots,\alpha_{rk(\mathfrak{g})}\}
\end{equation}
of $rk(\mathfrak{g})$ positive
elements of $R$ that generate all other positive elements as non-negative linear
combinations. These are the characteristics of simple roots; elements of $SR$ are 
``simple root junctions'' and elements of $R$ are ``root junctions''. Another 
non-trivial check is that
\begin{equation}
(A_\mathfrak{g})_{ij} = -(\alpha_i,\alpha_j)\, ,
\end{equation}
is the Cartan matrix of $\mathfrak{g}$\, .

To study more general representations, it is useful to have a map 
from junctions to their Dynkin labels
\begin{equation}
T: \bZ^N \to \bZ^{rk(\mathfrak{g})}\, ,
\end{equation}
where we remind the reader that the Dynkin labels are the basis of
$\bZ^{rk(\mathfrak{g})}$ in which
the simple roots of $\mathfrak{g}$ are represented by the associated row in the
Cartan matrix. Of course, $T$ is a matrix, and noting its definition and that
\begin{equation}
(A_\mathfrak{g})_{ij} = -(SR)^t \cdot I \cdot SR\, ,
\end{equation}
then we have
\begin{equation}
T = -(SR)^t \cdot I\, ,
\end{equation}
with $SR$ the $N\times rk(\mathfrak{g})$ matrix formed from the simple root 
junctions and $I$ the $N\times N$ matrix associated with the pairing $(\cdot\, ,\, \cdot)$.
Concretely,
\begin{equation}
I_{} = -\mathbb{1} + \frac{1}{2}(U + U^\text{t})\, ,
\end{equation}
where $U$ is an upper triangular matrix such that $U_{mn} = \gamma_m\cdot\gamma_n$
with $m<n$ and both in the set $\{1,\dots,N\}$.
This data will be computed explicitly in examples.

\vspace{.5cm}
String junctions in representations other than the adjoint can also be obtained.
In fact, string junctions may be realized for arbitrary Lie algebra representations \cite{DeWolfe:1998zf}, but we emphasize that this does not imply that all Lie algebra
representations are realized by string junctions in compact F-theory geometries.
For instance, symmetric tensor products arise in a particularly natural way \cite{Grassi:2013kha}.
For the purposes of this paper, it will suffice to study representations that arise via
a particularly simple method: we fix the asymptotic charge, and then find all
junctions $J$ with $(J,J)=-1$ that have that particular asymptotic charge. In particular, 
consider the Lie algebras arising from the Kodaira fibers considered in Section~\ref{sec:geom_monodromy},
with associated Lie algebras and ordered sets of vanishing
cycles:
\begin{center}
\begin{tabular}{c|c|c}
Fibration & Brane configuration & Algebra\\ \hline
$I_{4}$ & $\{1, 1, 1, 1\}$ & $\mathfrak{su}(4)$, $\mathfrak{sp}(2)$\\
$IV$ & $\{1, 3, 1, 3\}$& $\mathfrak{su}(3), \mathfrak{sp}(1)$\\
$IV^*$ & $\{1, 3, 1, 3, 1, 3, 1, 3\}$ & $\mathfrak{e}_6, \mathfrak{f}_4$\\
$I_{0}^*$ & $\{1, 3, 1, 3, 1, 3\}$ & $\mathfrak{so}(8), \mathfrak{so}(7), \mathfrak{g}_2$\\
$I_{1}^*$ & $\{1, 3, 1, 3, 1, 3, 1\}$& $\mathfrak{so}(10)$, $\mathfrak{so}(9)$
\end{tabular},
\end{center} 
One obtains the corresponding representations in Table~\ref{table:reps_and_charges}
by searching for all self-intersection $-1$ junctions with the asymptotic charges
listed in the Table~\ref{table:reps_and_charges}. The representation itself is determined 
by using the simple root junctions to determine the highest weight junction, and then
applying the Dynkin map. 

In summary, the data sufficient to determine the Lie algebra, including the set 
of roots $R$, is the ordered set of vanishing cycles $Z$, the pairing $(\cdot,\cdot)$,
and the notion of asymptotic charge.
This data arise naturally in the deformation, but we stress that the deformation is not necessary
if this data is otherwise available.

\vspace{1cm}
\noindent \emph{Group theoretical notations.}

In later discussions we will adopt the standard notation of labeling the names of the representations by the corresponding nodes of the Dynkin diagram. Although such notation is standard, since we will be using the non-standard Cartan matrices, it is worthwhile to explain it here. Note that the Cartan matrices that we will use in the next sections are related to the standard ones by transposing rows and columns of the matrices, which simply corresponds to relabeling the simple roots, so that the Cartan matrices we use are equivalent to the standard ones. To illustrate this, let us consider the example of $D_5$, corresponding to $SO(10)$.

\begin{figure*}[ht]
  \centering
  \includegraphics[width=.3\linewidth]{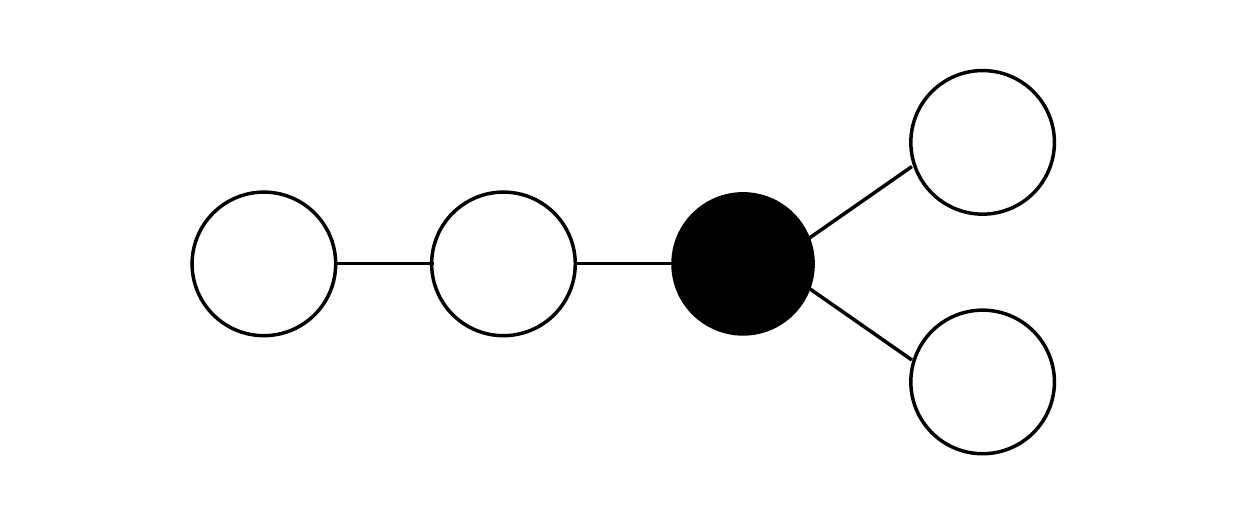}
  \caption{The central node of $D_5$}
  \label{fig:groupnotation_1}
\end{figure*}

The Cartan matrix we use for $SO(10)$ is
\[
\begin{pmatrix}
  -2 & 1 & 1 & 1 & 0 \\
  1 & -2 & 0 & 0 & 0 \\
  1 & 0 & -2 & 0 & 1 \\
  1 & 0 & 0 & -2 & 0 \\
  0 & 0 & 1 & 0 & -2
\end{pmatrix}.
\]
We can see that the simple root associated with the first row of the Cartan matrix corresponds to the central node of the $D_5$ Dynkin diagram as in Figure \ref{fig:groupnotation_1}.
In a similar vein, the simple roots associated with the second row and the fourth row of the Cartan matrix correspond to the upper-right and the lower-right node of the $D_5$ Dynkin diagram. We will make the choice that the the simple root associated with the second row of the Cartan matrix corresponds to the upper-right node of $D_5$. 

\begin{figure*}[ht]
  \centering
  \includegraphics[width=.3\linewidth]{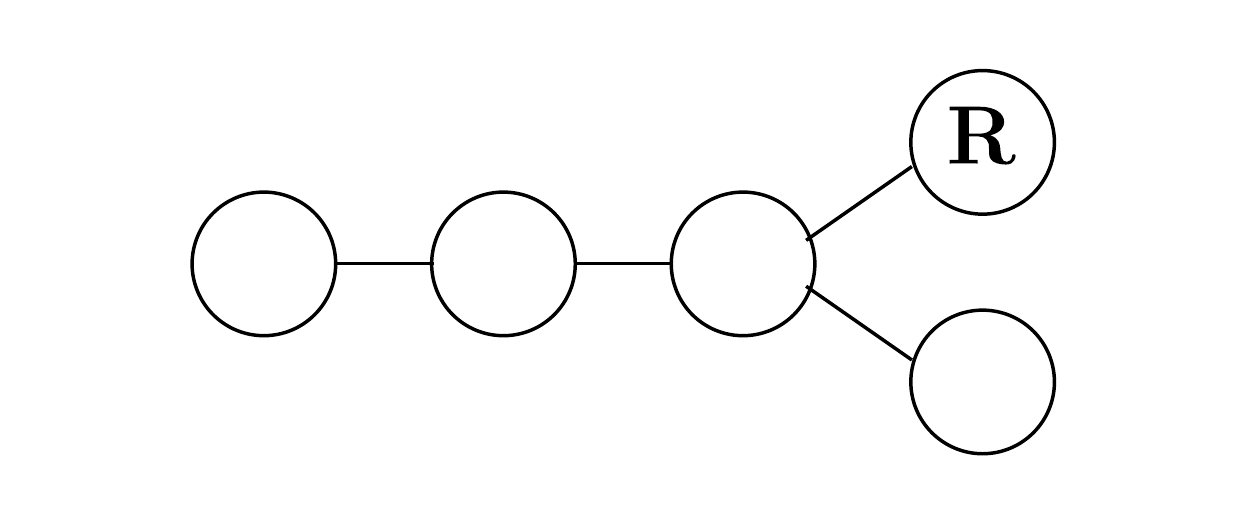}
  \caption{The representation $\mathbf{R}$ of $SO(10)$ with highest weight $[0,1,0,0,0]$}
  \label{fig:groupnotation_2}
\end{figure*}

There exists a representation $\mathbf{R}$ of $SO(10)$ with highest weight $[0,1,0,0,0]$. As this representation can be described with by a vector with a single entry of 1 at the second position of the weight vector and 0's otherwise, we can label $\mathbf{R}$ by node corresponding to the simple root associated with the second row the Cartan matrix which, as we have discussed above, is the upper-right node of $D_5$. This is shown in Figure \ref{fig:groupnotation_2}.

In this manuscript we will only be concerned with representations whose highest weight states take the form $[0,\dots,1,\dots,0]$, with only a single 1 in the weight vector which are known as fundamental representations. For a detailed introduction to this notation see \cite{ramond2010group}.

\begin{table}[ht]
\centering
\begin{tabular}{ c || c | c | c | c | c | c | c}
  Kodaira Fiber& \multicolumn{3}{c|}{$I_{0s}^*$} & $IV$ & $IV^*$ & $I_{1}^*$ & $I_{4}$ \\
  \hline
  Gauge group & \multicolumn{3}{c|}{$SO(8)$} & $SU(3)$ & $E_6$ & $SO(10)$ & $SU(4)$ \\
  \hline
  Asymp. Charge & $(1,0)$ & $(1,-1)$ & $(0,-1)$ & $(1,0)$ & $(1,0)$ & $(1,1)$ & $(1,0)$ \\
  \hline
  Representation & $\mathbf{8_v}$ & $\mathbf{8_s}$ & $\mathbf{8_c}$ & $\mathbf{3}$ & $\mathbf{27}$ & $\mathbf{16}$ & $\mathbf{4}$
\end{tabular}
\caption{Example representations obtained by searching for all junctions $J$ with the
given asymptotic charge and $(J,J)=-1$.}
\label{table:reps_and_charges}
\end{table}

\subsection{String Junctions on Singular Spaces}

We now introduce a formalism for studying string junctions on the singular space, which is relevant here because
the deformation that breaks the simply-laced algebra to the non-simply-laced algebra leaves the variety 
singular, and therefore derivation of the non-simply-laced algebra should be possible without a smoothing. 

Since the data of the Lie algebra is determined by the ordered set of vanishing cycles, the pairing,
and a notion of asymptotic charge, the relevant question is how to see each on the singular space. We will study the pairing in \cite{bcjt} and will instead 
focus on establishing ordered sets of vanishing cycles and asymptotic charge. We will use
a result from the mathematics literature to obtain a canonical ordered set of vanishing
cycles for each Kodaira fiber, and explain in a number of cases how they are related
to ordered sets obtained in the deformation picture.

First we establish a notion of asymptotic charge, which is
straightforward. Consider a singular elliptic fibration
\begin{equation}
X \xrightarrow{\pi} B\ , 
\end{equation}
that has a singular codimension one locus inside the discriminant, 
\begin{equation}
D \subset \Delta\, ,
\end{equation}
with Kodaira fiber $F=\pi^{-1}(p_D)$ for a generic point $p_D \in D$. In F-theory
language, there is a non-abelian seven-brane on $D$. Let $p\in B$ be a point near
a generic neighborhood of $D$, and as before define a reference elliptic fiber
\begin{equation}
E_p:=\pi^{-1}(p)\, .
\end{equation}
Let $C$ be any real curve from $p$ to a generic point $p_D\in D$. Some 
$\gamma\in H_1(E_p,\bZ)$ collapse upon following $C$ from $p$ to $p_D$,
defining a thimble or prong $\Gamma_C$. Then the asymptotic charge is $a(\Gamma_C)=\gamma$.

\vspace{.5cm}
Critical data associated the non-abelian seven-brane on $D$ is its Kodaira
fiber and associated monodromy $M_F\in SL(2,\bZ)$ and its multiplicity of vanishing in the
discriminant, $N:=mult_D(\Delta)$. Given that this data is central to the singular
elliptic fibration, it is a natural to ask whether there is a canonical way to 
associated a canonical ordered set of vanishing cycles with the pair $(M_F,N)$. 

\begin{table}
\centering
\begin{tabular}{c|c|c|c}
  Kodaira Fiber& $M_F$ & $N$ & Minimal Normal Factorization\\
  \hline
  $I_n$ & $\begin{pmatrix}1&n\\0&1\end{pmatrix}$ & $n$ & $M_1^n$ \\
  \hline
  $II$ & $\begin{pmatrix}0&1\\-1&1\end{pmatrix}$ & $2$ & $M_1M_3$ \\
  \hline
  $III$ & $\begin{pmatrix}0&1\\-1&0\end{pmatrix}$ & $3$ & $M_1M_3M_1$ \\
  \hline
  $IV$ & $\begin{pmatrix}-1&1\\-1&0\end{pmatrix}$ & $4$ & $(M_1M_3)^2$ \\
  \hline
  $I_n^*$ & $\begin{pmatrix}-1&-n\\0&-1\end{pmatrix}$ & $n+6$ & $M_1^n(M_1M_3)^3\ (=-M_1^n)$ \\
  \hline
  $IV^*$ & $\begin{pmatrix}0&-1\\1&-1\end{pmatrix}$ & $8$ & $M_1M_3(M_1M_3)^3\ (=-M_1M_3)$ \\
  \hline
  $III^*$ & $\begin{pmatrix}0&-1\\1&0\end{pmatrix}$ & $9$ & $M_1M_3M_1(M_1M_3)^3\ (=-M_1M_3M_1)$ \\
  \hline
  $II^*$ & $\begin{pmatrix}1&-1\\1&0\end{pmatrix}$ & $10$ & $(M_1M_3)^2(M_1M_3)^3\ (=-(M_1M_3)^2)$
\end{tabular}
\caption{Kodaira fibers and their monodromy matrix and minimal normal
factorization.}
\label{table:mnf}
\end{table}

To do so, we will utilize results of \cite{cadavid2009normal}, which we now briefly review.
Two particular $SL(2,\bZ)$ matrices are central to the results, which in \cite{cadavid2009normal} are called
\begin{equation}
  U = 
  \begin{pmatrix}
    1 & 1 \\
    0 & 1
  \end{pmatrix},\qquad
  V =
  \begin{pmatrix}
    1 & 0 \\
    -1 & 1
  \end{pmatrix}.
\end{equation}
In the notation that will appear later\footnote{The use of the $M_i$ is standard notation in the string junction literature cited throughout this manuscript.}, we have
\begin{equation}
  U = M_1, \qquad V = M_3,
\end{equation}
where
\begin{equation}
  M_1 = 
  \begin{pmatrix}
    1 & 1 \\
    0 & 1
  \end{pmatrix},\qquad
  M_2 = 
  \begin{pmatrix}
    0 & 1 \\
    -1 & 2
  \end{pmatrix},\qquad
  M_3 =
  \begin{pmatrix}
    1 & 0 \\
    -1 & 1
  \end{pmatrix}.
\end{equation}
The monodromy matrix $M_F$ admits a factorization
into
\begin{align}
  M_F = G_1\cdot G_2\cdots G_N,
\end{align}
where each $G_i$ is the monodromy matrix that is associated with
a $(p,q)$ seven-brane, that is, it is of the form:
\begin{equation}
	M_{p,q} =
	\begin{pmatrix}
    1-pq & p^2 \\
    -q^2 & 1+pq
  	\end{pmatrix}.
\end{equation}
However, there are different possible ordered sets of vanishing cycles, which would give rise to different factorizations. The so-called \emph{minimal normal factorizations} are presented in Table \ref{table:mnf}.
It is important to note that this minimal normal factorization only exists for the $SL(2,\mathbb{Z})$ matrices associated with the Kodaira fiber types, and does not exist for general $SL(2,\mathbb{Z})$ matrices. Different factorizations are related to one another by so-called \emph{Hurwitz moves}.

\vspace{.3cm}
\noindent \emph{{\bf Definition.} Let $G$ be a group and let $g_1\dots g_k$ be products of elements of $G$. Another such product $g_1'\dots g_k'$ is said to be obtained from $g_1\dots g_k$ via a Hurwitz move if for some $1\leq i\leq k-1$, $g_j'\neq g_j$ for $j\notin \{i, i+1\}$ and either $g_i'\neq g_{i+1}$, $g_{i+1}'=g_{i+1}^{-1}g_i g_{i+1}$ or $g_i'=g_i g_{i+1} g_i^{-1}$, $g_{i+1}'=g_i$. We will also say that an ordered set $\{g_1', g_2', \dots, g_k'\}$ is obtained from another ordered set $\{g_1, g_2, \dots, g_k\}$ by applying one Hurwitz move, if the same relations hold between $g_i'$'s and the $g_i$'s. 
}

\vspace{.3cm}
\noindent That is, a Hurwitz move does a transformation of the form
\begin{equation}
  g_1\dots g_i g_{i+1}\dots g_k \rightarrow g_1\dots g_{i+1}(g_{i+1}^{-1}g_i g_{i+1})\dots g_k,
\end{equation}
or of the form
\begin{equation}
  g_1\dots g_i g_{i+1}\dots g_k \rightarrow g_1\dots (g_i g_{i+1} g_i^{-1})g_i\dots g_k,
\end{equation}
i.e., $g_{i+1}$ is ``pulled past" $g_i$, conjugating it in the process, or vice versa\footnote{In a deformation picture, Hurwitz moves can arise naturally via brane rearrangement or choosing different paths to $I_1$ fibers. Both induce Hanany-Witten moves on the junction basis.}. This purely algebraic definition makes sense for the monodromy on the singular space.

One theorem of \cite{cadavid2009normal} will be critical for us, where the possible factorization fall under two different cases: 

\vspace{.3cm}
\noindent \emph{{\bf Theorem.} Let $M$ be a matrix that corresponds to the monodromy
of a singular fiber in an elliptic fibration. If $M=G_1\dots G_r$ is a factorization
of $M$ in terms of conjugates of $U$ (i.e., in terms of $M_{(p,q)}$-type matrices), then $r$ is greater than or equal to $n$, the number of factors in the m.n.f. of $M$. After a finite number of Hurwitz moves it is possible to obtain: 
\begin{itemize}
\item Cases $wI_n-IV$: $G_1\dots G_r=C_1\dots C_n(VU)^{6s}$, with $C_1\dots C_n$ the m.n.f. of $M$ and $s=(r-n)/12$.
\item Cases $I_n^*-IV^*$: $G_1\dots G_r=C_1\dots C_n(VU)^{6s+3}$, with $C_1\dots C_n$ the m.n.f. of $-M$ and  $s=(r-n-6)/12$.
\end{itemize} }

\vspace{.3cm}
\noindent 
This theorem is essential for us, because it means that, given
a pair $(M_F,N)$ and up to Hurwitz moves,
we can canonically choose the ordered set of vanishing cycles associated with the minimal
normal factorization, and we can do this on the singular space; in doing
so, we are automatically considering the case $r=n$. We will also take the
associated pairing, and in \cite{bcjt} we will show that the pairing is well-behaved
under Hurwitz moves. With this data motivated on the singular space, we may perform
calculations there, as well. This approach will be further justified because
the new calculations in F-theory agree with the conclusions drawn from the
M-theory Coulomb branch.

It is also worth noting that the ordered set of vanishing cycles associated with
the minimal normal factorization is in many cases equivalent to the ones obtained
by simple deformations and following straight line paths to $I_1$ fibers, see, e.g., \cite{Grassi:2013kha,Grassi:2014sda}.

\subsection{Automorphisms and non-simply-laced algebras in examples}\label{sec:cases_analytic}

In this section we demonstrate the monodromy reduction of string junction states under deformation of a 7-brane fiber from split to non-split. We begin with the case of $I_{0s}^*$: $SO(8)\rightarrow G_2$.
\subsubsection{Type $I_{0s}^*$: $SO(8)\rightarrow G_2$}\label{sec:G2_junc_analytic}

We begin by analyzing the case of $SO(8)$ breaking to $G_2$.
The geometric monodromy action is analyzed in Section \ref{sec:G2_analytic}.  Let us present data relevant to junctions in representations of $SO(8)$. 
The ordered set of seven branes at $z = 0$ can be chosen to be
\begin{equation}
Z=\{1,3,1,3,1,3\}.
\end{equation} The intersection matrix is:
\begin{center}
  $I = 
  \begin{pmatrix}
    -1 & 1/2 & 0 & 1/2 & 0 & 1/2 \\
    1/2 & -1 & -1/2 & 0 & -1/2 & 0 \\
    0 & -1/2 & -1 & 1/2 & 0 & 1/2 \\
    1/2 & 0 & 1/2 & -1 & -1/2 & 0 \\
    0 & -1/2 & 0 & -1/2 & -1 & 1/2 \\
    1/2 & 0 & 1/2 & 0 & 1/2 & -1
  \end{pmatrix}.$
\end{center}
One choice of simple roots in this junction basis are:
\begin{align*}
  \alpha_1 &= (0, 0, 0, 1, 0, -1) \\
  \alpha_2 &= (0, 0, 1, 0, -1, 0) \\
  \alpha_3 &= (0, 1, -1, -1, 1, 0) \\
  \alpha_4 &= (1, 0, -1, 0, 0, 0).
\end{align*}
Direct computation gives the Cartan matrix
\begin{equation}
\begin{pmatrix}
-2 & 1 & 0 & 0 \\
1 & -2 & 1 & 1 \\
0 & 1 & -2 & 0 \\
0 & 1 & 0 & -2
\end{pmatrix},
\end{equation}
and the Dynkin map
\[
T
= 
\begin{pmatrix}
  0 & 0 & 0 & 1 & 1 & -1 \\
  0 & 0 & 1 & -1 & -1 & 0 \\
  0 & 1 & 0 & 0 & 1 & 0 \\
  1 & -1 & -1 & 0 & 0 & 0
\end{pmatrix}.
\]
This matrix maps weight junctions to their Dynkin labels. By choosing 
an asymptotic charge, finding all junctions of self-intersection $-1$ with that asymptotic
charge, and using the roots to find the highest weight, we may find certain representations
of $SO(8)$. 
From appendix \ref{sec:G2_junc_comp}, we recall highest weight junctions of various representations and asymptotic charges that will be important for us. They are: 
\begin{center}
\begin{tabular}{c|c|c|c}
Asymptotic Charge & Highest Weight Junction & Dynkin Label & Representation\\
\hline
$(1,0)$ & $(1, 0, 0, 0, 0, 0)$ & \multirow{3}{*}{$[0,0,0,1]$} &\multirow{3}{*}{$\mathbf{8_v}$} \\
$(1,-2)$ & $(0, -1, 0, -1, 1, 0)$& \\
$(-1,-2)$ & $(0, 0, -1, -1, 0, -1)$& \\
\hline
$(1,-1)$ & $(0, 0, 0, -1, 1, 0)$ & \multirow{3}{*}{$[0,0,1,0]$} & \multirow{3}{*}{$\mathbf{8_s}$}   \\
$(-1,-1)$ & $(0, 1, -1, -1, 0, -1)$ &\\
$(1,1)$ & $(1, 1, 0, 0, 0, 0)$ & \\
\hline
$(0,-1)$ & $(0, 0, 0, 0, 0, -1)$ & $[1,0,0,0]$ & $\mathbf{8_c}$
\end{tabular}
\end{center}
We have identified the representations according to their Dynkin labels. Of course more representations
exist, including eight-dimensional representations with different asymptotic
charges, but we have listed the data that will be relevant for our monodromy calculations.

We now turn to the monodromy action on string junctions. Recall from Section \ref{sec:G2_analytic} that upon turning on the deformation $\epsilon\neq 0$, the three loci where the $I_1$ locus intersects the $I_0^*$ locus split into three pairs of roots, and we computed the geometric monodromy associated to each of the three pairs. We found one was a double rotation of $x_1$-$x_2$ , another was a double rotation of $x_1$-$x_3$ , and another was a double rotation of $x_2$-$x_3$. Such rotations are realized as braidings in the geometry, in the sense of \cite{Grassi:2016bhs}. However, the deformation splits the points in the pair, and we may also take a loop around one of them in each pair, which induces the monodromies
\begin{equation}
M_1 = 
\begin{pmatrix}
	1 & 1\\0 & 1
\end{pmatrix}, \quad 
M_2 = 
\begin{pmatrix}
	0 & 1\\-1 & 2
\end{pmatrix}, \quad
M_3 = 
\begin{pmatrix}
	1 & 0\\-1 & 1
\end{pmatrix}.
\label{eq:monodromy_mats}
\end{equation}
Each of these loops around one of the points in each pair may be traversed by a D3-brane, and we will refer to them as loop $1$, loop $2$, and loop $3$, respectively. This monodromy behavior persists in certain deformations to Weierstrass 
models that do not have non-isolated singularities \cite{Grassi:toappear}.

Let us traverse loop $3$. Its monodromy, $M_3$, induces a map on asymptotic charge as
\begin{align}
\begin{pmatrix}1 \\ 0\end{pmatrix} \qquad &\mapsto \qquad
\begin{pmatrix}1 \\ -1\end{pmatrix}\, ,
\label{eq:8v_8s}
\end{align} 
which we see corresponds to a map on
representations 
\begin{equation}
\mathbf{8_v} \qquad \mapsto \qquad \mathbf{8_s}\, .
\label{eq:8v_8s1}
\end{equation}
Repeating the loop a second time maps $\bf{8}_s$ back to $\bf{8}_v$, but with asymptotic charge $(1,-2)$. Traversing loop $2$ induces a map on asymptotic charge as 
\begin{align}
\begin{pmatrix}1 \\ 0\end{pmatrix} \qquad &\mapsto \qquad
\begin{pmatrix}0 \\ -1\end{pmatrix}\, ,
\label{eq:8v_8c}
\end{align} 
which corresponds to a map on representations
\begin{equation}
\mathbf{8_v} \qquad \mapsto \qquad \mathbf{8_c}\, .
\label{eq:8v_8c1}
\end{equation}
Traversing the loop a second time transforms it back to $\mathbf{8_v}$ with asymptotic charge $(-1,-2)$. Similarly, traversing loop one maps
\begin{align}
\begin{pmatrix}0 \\ -1\end{pmatrix} \qquad &\mapsto \qquad
\begin{pmatrix}-1 \\ -1\end{pmatrix}
\label{eq:8c_8s}
\end{align} 
which maps the representation as
\begin{equation}
\mathbf{8_c} \qquad \mapsto \qquad \mathbf{8_s},
\label{eq:8c_8s1}
\end{equation}
and a second traversal maps it back to $\mathbf{8_c}$, but with asymptotic charge $(0,-1)$.

\begin{figure*}[ht]
  \centering
  \includegraphics[width=.5\linewidth]{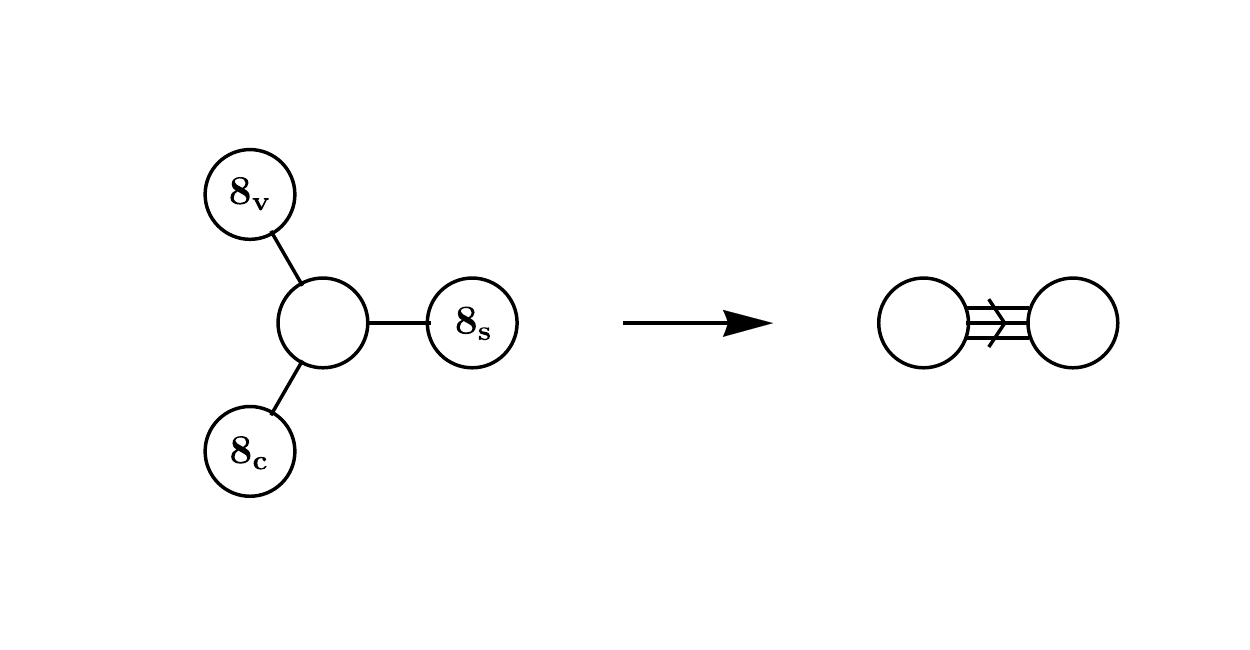}
  \caption{The outer-automorphism acting on $D_4$ diagram of $\mathfrak{so}_8$ leads to $G_2$ diagram of $\mathfrak{g}_2$.}
  \label{fig:I0*ns_dyn_mono}
\end{figure*}

These loops can be taken arbitrarily small, and as argued we should therefore identify the associated states. The Dynkin labels of the highest weights of the three eight-dimensional
representations each mark one of the exterior node of the Dynkin diagram, i.e., the
node corresponding to the placement of its non-zero entry.
This fact, together with the monodromy action that we have derived,
shows that the combined set of monodromies around the three loops
gives rise to an  $S_3$ outer-automorphism acting on $D_4$ which, after quotienting, gives rise to $G_2$. This can be seen from the Dynkin diagram as identifying all the three nodes, as shown in Figure~\ref{fig:I0*ns_dyn_mono}.

\vspace{.5cm}
\textit{Remark.} The same kind of argument can be applied to all the cases discussed in Section \ref{sec:cases_analytic}. The key is to identify the $SL(2, \mathbb{Z})$ matrices acting on the asymptotic charges corresponding to the geometric monodromy actions. The reduced gauge group and the matter representation after the identification therefore follow naturally from the the structure of the Dynkin digram of the relevant simply-laced Lie algebra, and the outer-automorphisms acting on it.

\subsubsection{Type $I_{0ss}^*$: $SO(8)\rightarrow SO(7)$}\label{sec:SO7_junc_analytic}

The geometric monodromy action is analyzed in Section \ref{sec:SO7analytic}. The data associated to $SO(8)$, including
the simple roots, junction-to-Dynkin map,
and correspondence between $(p,q)$ charges
and representations are the same as in
Section \ref{sec:G2_junc_analytic} (see also Appendix \ref{sec:G2_junc_comp} and \ref{sec:SO7_junc_comp}.). 

\begin{figure*}[ht]
  \centering
  \includegraphics[width=.5\linewidth]{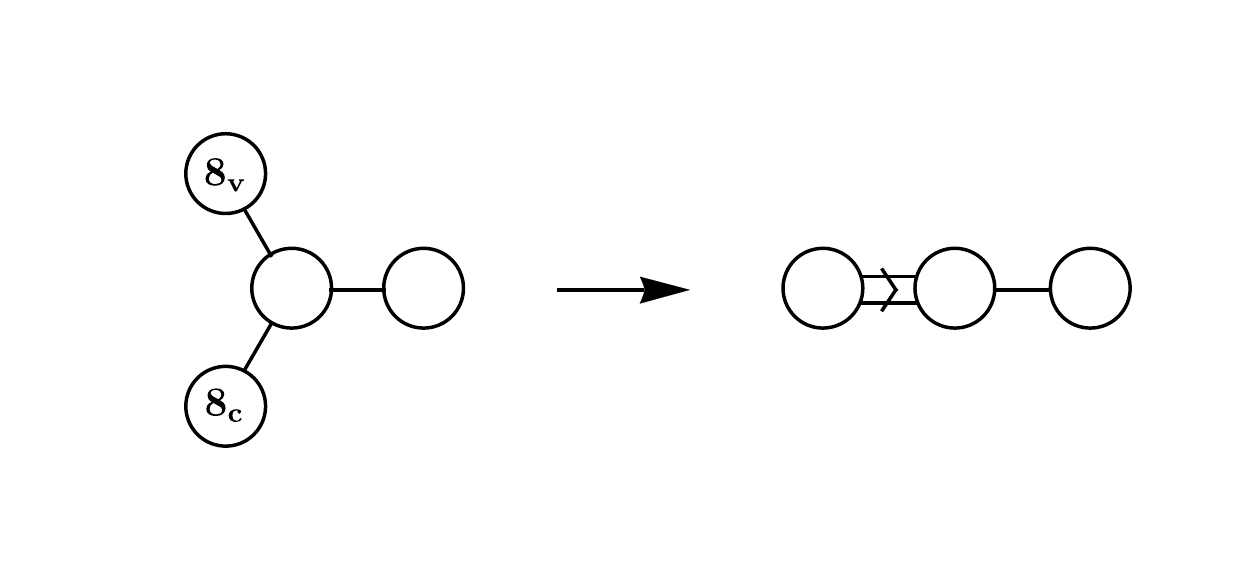}
  \caption{The outer-automorphism acting on the $D_4$ diagram of $\mathfrak{so}_8$ leads to the $B_3$ diagram of $\mathfrak{so}_7$.}
  \label{fig:I0*ss_dyn_mono}
\end{figure*}

The central difference between the 
$SO(8)\to SO(7)$ breaking that we study
here and the $SO(8)\to G_2$ breaking
of Section~\ref{sec:G2_junc_analytic} lies in the different pattern of the splitting of the roots. In Section~\ref{sec:SO7analytic} we showed that only a single new loop arises upon
deformation from $SO(8)\to SO(7)$, and accordingly
only a single new monodromy action can arise. As shown in
section \ref{sec:SO7analytic}, the monodromy
associated with the loop that appears is
$M_3$. It induces a map on $(1,0)$ asymptotic
charge given by
\begin{align}
\begin{pmatrix}1 \\ 0\end{pmatrix} \qquad &\mapsto \qquad
\begin{pmatrix}1 \\ -1\end{pmatrix}
\end{align} 
which we see corresponds to a map on
representations
\begin{equation}
\mathbf{8_v} \qquad \mapsto \qquad \mathbf{8_s}.
\end{equation}
Traversing the loop a second time maps back to $\mathbf{8_v}$. This gives rise to a $\mathbb{Z}_2$ outer-automorphism $D_4$ that acts on the Dynkin diagram as shown in Figure~\ref{fig:I0*ss_dyn_mono}. This identifies two of the three nodes, labeled by $\mathbf{8_v}$ and $\mathbf{8_c}$. Quotienting by this automorphism reduces the algebra to $SO(7)$. The detailed correspondence between the relevant string junctions with given asymptotic charges and matter representations is in Appendix \ref{sec:G2_junc_comp} and \ref{sec:SO7_junc_comp}.

\subsubsection{Type $IV$: $SU(3)\rightarrow Sp(1)$}\label{sec:IV_junc_analytic}

The geometric monodromy action relevant for the reduction $SU(3)\rightarrow Sp(1)$ was analyzed in Section \ref{sec:IVanalytic}. The ordered set of seven branes at $z = 0$ can be taken to be
\begin{equation}
Z = \{1,3,1,3\},
\end{equation}
in which case the 
 intersection matrix takes the form:
\begin{center}
  $I = 
  \begin{pmatrix}
    -1 & 1/2 & 0 & 1/2 \\
    1/2 & -1 & -1/2 & 0 \\
    0 & -1/2 & -1 & 1/2 \\
    1/2 & 0 & 1/2 & -1
  \end{pmatrix}$.
\end{center}
One choice of simple roots in the junction basis are:
\begin{align*}
  \alpha_1 &= (0, 1, 0, -1) \\
  \alpha_2 &= (1, 0, -1, 0)\, ,
\end{align*}
and the Cartan matrix is:
\[
\begin{pmatrix}
    -2 & 1 \\
    1 & -2  \\
\end{pmatrix}\,
\]
and the Dynkin map is:
\[
T
= 
\begin{pmatrix}
    0 & 1 & 1 & -1\\
    1 & -1 & -1 & 0 \\
\end{pmatrix}\, .
\]

The highest weight junctions of the various representations and asymptotic charges that will be important for us are:
\begin{center}
\begin{tabular}{c|c|c|c}
Asymptotic Charge & Highest Weight Junction & Dynkin Label & Representation \\
\hline
$(1,0)$ & $(1,0,0,0)$ & \multirow{2}{*}{$[0,1]$} & \multirow{2}{*}{$\mathbf{3}$} \\
$(-1,-1)$ & $(0,0,-1,-1)$ & \\
\hline
$(0,-1)$ & $(0,0,0,-1)$ & $[1,0]$ & $\mathbf{\bar{3}}$
\end{tabular}
\end{center}
We recognize that $[0,1]$ is the highest weight state of $\mathbf{3}$ and $[1,0]$ the highest weight state of $\mathbf{\bar{3}}$ (see Appendix~\ref{sec:IV_junc_comp} for further details).

Determining the associated $SL(2,\bZ)$ monodromy
is somewhat more involved. After turning on $\epsilon$, type $IV_s$ becomes type $IV_{ns}$. Recall that in Section \ref{sec:IVanalytic} we showed that the monodromy action induced by looping around one of the pairs of the splitting roots of $\Delta_R(t; \delta; \epsilon) = 0$, which corresponds to an $x_1$-$x_2$ swap, followed by an $x_1$-$x_3$ swap. 

There is a slight ambiguity here; a priori it is not clear if the total monodromy action upon encircling the $I_1$ should be $M_1 \cdot M_3$ or $M_3 \cdot M_1$. However, this can be fixed by the observation that the monodromy around non-abelian 7-brane is twice the monodromy around the $I_1$ locus, which can be shown explicitly. In this basis of vanishing cycles, the monodromy around the non-abelian 7-brane is $(M_1 \cdot M_3)^2$, and we therefore conclude the $I_1$ monodromy is $M_1 \cdot M_3$. 

\begin{figure*}[ht]
  \centering
  \includegraphics[width=.5\linewidth]{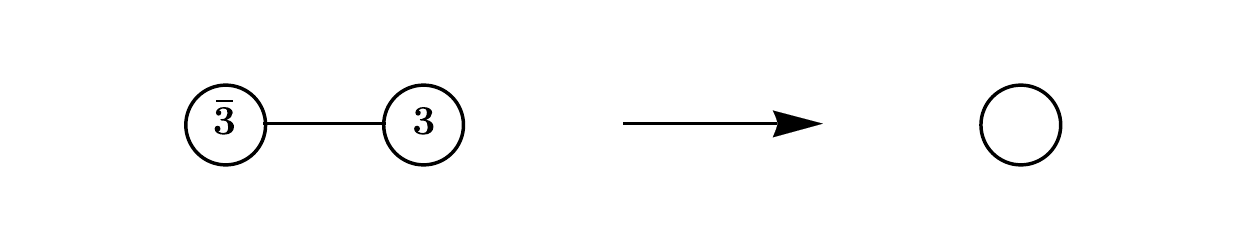}
  \caption{The outer-automorphism acting on $A_2$ diagram of $\mathfrak{su}(3)$ leads to $A_1$ diagram of $\mathfrak{su}(2)$.}
  \label{fig:IV_dyn_mono}
\end{figure*}

The $I_1$ monodromy
\begin{align}
	M_1\cdot M_3 = 
	\begin{pmatrix}
		0 & 1 \\ -1 & 1
	\end{pmatrix} \, ,
\end{align}
gives the transformation on the asymptotic charge
\begin{align}
	\begin{pmatrix}
		1 \\ 0
	\end{pmatrix}
	\qquad \mapsto \qquad
	\begin{pmatrix}
		0 \\ -1
	\end{pmatrix}.
\end{align}
This monodromy action corresponds to a map on representations $\mathbf{3}\mapsto\mathbf{\bar{3}}$. Traversing the loop a second time induces another $M_1\cdot M_3$ action, which maps the asymptotic charge to $(-1,-1)$, and brings us back to the $\mathbf{3}$ representation. This gives rise to an $\mathbb{Z}_2$ outer-automorphism of $A_2$ identifying the two nodes therefore leads to the $\mathfrak{su}(2)$ $A_1$ Dynkin diagram as is shown in Fig.\ref{fig:IV_dyn_mono}. That is, turning on the deformation $\epsilon \neq 0$ reduces the symmetry algebra to $A_1$.

\subsubsection{Type $IV^*$: $E_6\rightarrow F_4$}\label{sec:IV*_junc_analytic}

The ordered set of seven branes at $z = 0$ can be taken to be $\{1,3,1,3,1,3,1,3\}$. The intersection matrix is:
\begin{center}
  $I =
  \begin{pmatrix}
    -1 & 1/2 & 0 & 1/2 & 0 & 1/2 & 0 & 1/2 \\
    1/2 & -1 & -1/2 & 0 & -1/2 & 0 & -1/2 & 0 \\
    0 & -1/2 & -1 & 1/2 & 0 & 1/2 & 0 & 1/2 \\
    1/2 & 0 & 1/2 & -1 & -1/2 & 0 & -1/2 & 0 \\
    0 & -1/2 & 0 & -1/2 & -1 & 1/2 & 0 & 1/2 \\
    1/2 & 0 & 1/2 & 0 & 1/2 & -1 & -1/2 & 0 \\
    0 & -1/2 & 0 & -1/2 & 0 & -1/2 & -1 & 1/2 \\
    1/2 & 0 & 1/2 & 0 & 1/2 & 0 & 1/2 & -1
  \end{pmatrix} 
  $.
\end{center}
The simple roots in junction basis are:
\begin{align*}
  \alpha_1 &= (0, 0, 0, 0, 0, 1, 0, -1) \\
  \alpha_2 &= (0, 0, 0, 0, 1, 0, -1, 0) \\
  \alpha_3 &= (0, 0, 0, 1, -1, -1, 1, 0) \\
  \alpha_4 &= (0, 0, 1, 0, -1, 0, 0, 0) \\
  \alpha_5 &= (0, 1, -1, -1, 0, -1, 1, 1) \\
  \alpha_6 &= (1, 0, -1, 0, 0, 0, 0, 0),
\end{align*}
and the Cartan matrix is:
\[
\begin{pmatrix}
    -2 & 1 & 0 & 0 & 1 & 0 \\
    1 & -2 & 1 & 1 & 0 & 0 \\
    0 & 1 & -2 & 0 & 0 & 0 \\
    0 & 1 & 0 & -2 & 0 & 1 \\
    1 & 0 & 0 & 0 & -2 & 0 \\
    0 & 0 & 0 & 1 & 0 & -2
\end{pmatrix}.
\]
and the junction-to-Dynkin map is:
\[
T
= 
\begin{pmatrix}
    0 & 0 & 0 & 0 & 0 & 1 & 1 & -1 \\
    0 & 0 & 0 & 0 & 1 & -1 & -1 & 0 \\
    0 & 0 & 0 & 1 & 0 & 0 & 1 & 0 \\
    0 & 0 & 1 & -1 & -1 & 0 & 0 & 0 \\
    0 & 1 & 0 & 0 & 0 & 0 & 0 & 1 \\
    1 & -1 & -1 & 0 & 0 & 0 & 0 & 0
\end{pmatrix}.
\]
The highest weight junctions of various representations and asymptotic charges that will be important for us take for form:
\begin{center}
\begin{tabular}{c|c|c|c}
Asymptotic Charge & Highest Weight Junction & Dynkin Label & Representation \\
\hline
$(1,0)$ & $(1,0,0,0,0,0,0,0)$ & \multirow{2}{*}{$[0,0,0,0,0,1]$} & \multirow{2}{*}{$\mathbf{27}$} \\
$(-1,-1)$ & $(1,1,-1,0,-1,-1,0,-1)$ & & \\
\hline
$(0,-1)$ & $(0,1,-1,-1,0,-1,1,0)$ & $[0,0,0,0,1,0]$ & $\mathbf{\overline{27}}$
\end{tabular}
\end{center}
We recognize that $[0,0,0,0,0,1]$ as the highest weight state of $\mathbf{27}$ and $[0,0,0,0,1,0]$ as the highest weight state of $\mathbf{\overline{27}}$. For details see Appendix \ref{sec:IV*_junc_comp}.

\begin{figure*}[ht]
  \centering
  \includegraphics[width=.5\linewidth]{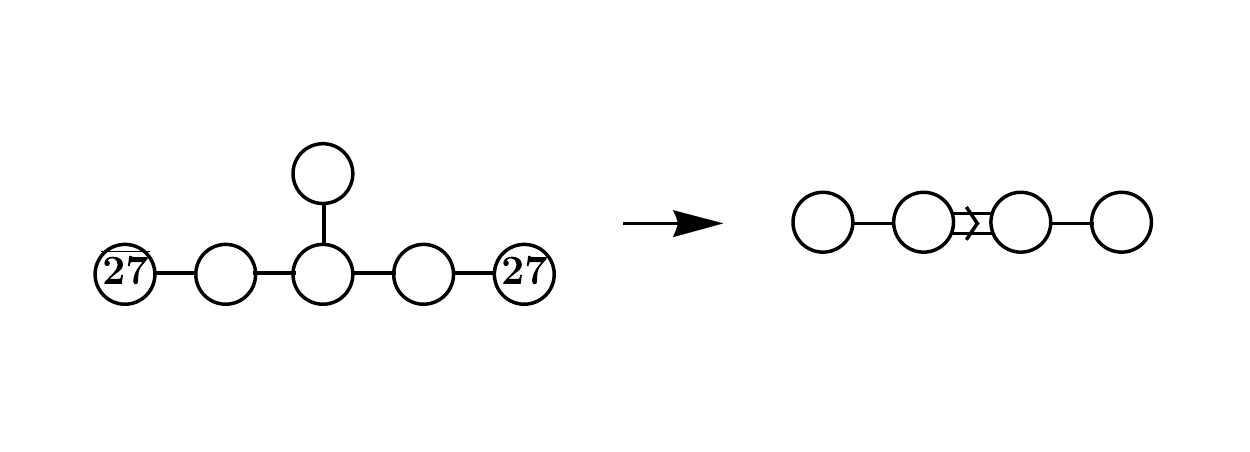}
  \caption{The outer-automorphism acting on $E_6$ diagram of $\mathfrak{e}_6$ leads to $F_4$ diagram of $\mathfrak{f}_4$.}
  \label{fig:IV*_dyn_mono}
\end{figure*}

As derived in section \ref{sec:IV*analytic},
the $SL(2,\mathbb{Z})$ matrix corresponding to the geometric monodromy action is $M_1\cdot M_3$. As the D3-brane traverses the loop, the $M_1\cdot M_3$ action induces
\begin{equation}
	\begin{pmatrix}
		1 \\ 0
	\end{pmatrix}
	\qquad \mapsto \qquad
	\begin{pmatrix}
		0 \\ -1
	\end{pmatrix}
\end{equation}
which corresponds to a map on representations:
\begin{equation}
	\mathbf{27}\qquad\mapsto\qquad\mathbf{\overline{27}}.
\end{equation}
A subsequent $M_1\cdot M_3$ action from traversing the loop around the other pair of roots induces
\begin{equation}
	\begin{pmatrix}
		0 \\ -1
	\end{pmatrix}
	\qquad \mapsto \qquad
	\begin{pmatrix}
		-1 \\ -1
	\end{pmatrix}
\end{equation}
which corresponds to a map on representations:
\begin{equation}
	\mathbf{\overline{27}}\qquad\mapsto\qquad\mathbf{27}.
\end{equation}
This monodromy action swaps the representations $\mathbf{27}$ and $\mathbf{\overline{27}}$, giving rise to
a $\mathbb{Z}_2$ outer-automorphism acting on $E_6$ Dynkin diagram, as shown in Figure~\ref{fig:IV*_dyn_mono}. Quotienting by it gives $F_4$.

\subsubsection{Type $I_1^*$: $SO(10)\rightarrow SO(9)$}\label{sec:I1*analytic}

The geometric monodromy action was analyzed in Section \ref{sec:I1*_junc_analytic}. The ordered set of seven branes at $z = 0$ is $\{1,3,1,3,1,3,1\}$. The intersection matrix is:
\begin{center}
  $I = 
  \begin{pmatrix}
    -1 & 1/2 & 0 & 1/2 & 0 & 1/2 & 0 \\
    1/2 & -1 & -1/2 & 0 & -1/2 & 0 & -1/2 \\
    0 & -1/2 & -1 & 1/2 & 0 & 1/2 & 0 \\
    1/2 & 0 & 1/2 & -1 & -1/2 & 0 & -1/2 \\
    0 & -1/2 & 0 & -1/2 & -1 & 1/2 & 0 \\
    1/2 & 0 & 1/2 & 0 & 1/2 & -1 & -1/2 \\
    0 & -1/2 & 0 & -1/2 & 0 & -1/2 & -1
  \end{pmatrix}$.
\end{center}
The simple roots in junction basis are:
\begin{align*}
  \alpha_1 &= (0, 0, 0, 0, 1, 0, -1) \\
  \alpha_2 &= (0, 0, 0, 1, -1, -1, 1) \\
  \alpha_3 &= (0, 0, 1, 0, -1, 0, 0) \\
  \alpha_4 &= (0, 1, -1, -1, 0, 0, 1) \\
  \alpha_5 &= (1, 0, -1, 0, 0, 0, 0)\, .
\end{align*}
The Cartan matrix is:
\[
\begin{pmatrix}
  -2 & 1 & 1 & 1 & 0 \\
  1 & -2 & 0 & 0 & 0 \\
  1 & 0 & -2 & 0 & 1 \\
  1 & 0 & 0 & -2 & 0 \\
  0 & 0 & 1 & 0 & -2
\end{pmatrix}.
\]
and the junction-to-Dynkin map is:
\[
T
= 
\begin{pmatrix}
  0 & 0 & 0 & 0 & 1 & -1 & -1 \\
  0 & 0 & 0 & 1 & 0 & 0 & 1 \\
  0 & 0 & 1 & -1 & -1 & 0 & 0 \\
  0 & 1 & 0 & 0 & 0 & 1 & 1 \\
  1 & -1 & -1 & 0 & 0 & 0 & 0
\end{pmatrix}.
\]

The highest weight junctions of various representations and asymptotic charges that will be important for us are:
\begin{center}
\begin{tabular}{c|c|c|c}
Asymptotic Charge & Highest Weight Junction & Dynkin Label & Representation \\
\hline
$(1,1)$ & $(1,1,0,0,0,0,0)$ & \multirow{2}{*}{$[0,0,0,1,0]$} & \multirow{2}{*}{$\mathbf{16}$} \\
$(3,1)$ & $(1,0,1,0,1,1,0)$ & \\
\hline
$(2,1)$ & $(1,0,1,1,0,0,0)$ & $[0,1,0,0,0]$ & $\mathbf{\overline{16}}$
\end{tabular}
\end{center}
We recognize that $[0,0,0,1,0]$ is the highest weight state of $\mathbf{16}$ and $[0,1,0,0,0]$ the highest weight state of $\mathbf{\overline{16}}$. See Appendix \ref{sec:I1*_junc_comp} for details.

In this case we can simply read off the monodromy matrix as $M_1$, as the geometric monodromy is a simple $x_1$-$x_3$ swap, Looping around the first root of $\Delta_R(t; \delta; \epsilon) = 0$ followed by looping around the second, the $M_1$ matrix acts on the asymptotic charge as:
\begin{equation}
	\begin{pmatrix}
		1 \\ 1
	\end{pmatrix}
	\qquad \mapsto \qquad
	\begin{pmatrix}
		2 \\ 1
	\end{pmatrix}
\end{equation}
which corresponds to a map on representations:
\begin{equation}
	\mathbf{16}\qquad\mapsto\qquad\mathbf{\overline{16}}.
\end{equation}
Traversing the loop around the other root in the pair of the splitting roots induces another $M_1$ action on the asymptotic charge:
\begin{equation}
	\begin{pmatrix}
		2 \\ 1
	\end{pmatrix}
	\qquad \mapsto \qquad
	\begin{pmatrix}
		3 \\ 1
	\end{pmatrix}
\end{equation}
which corresponds to a map on representations:
\begin{equation}
	\mathbf{\overline{16}}\qquad\mapsto\qquad\mathbf{16}.
\end{equation}
Hence the representations $\mathbf{16}$ and $\mathbf{\overline{16}}$ of $SO(10)$ are identified under this action. This identification gives rise to a $\mathbb{Z}_2$ outer-automorphism acting on $D_5$ Dynkin diagram shown in Figure~{\ref{fig:I1*_dyn_mono}:

\begin{figure*}[ht]
  \centering
  \includegraphics[width=.5\linewidth]{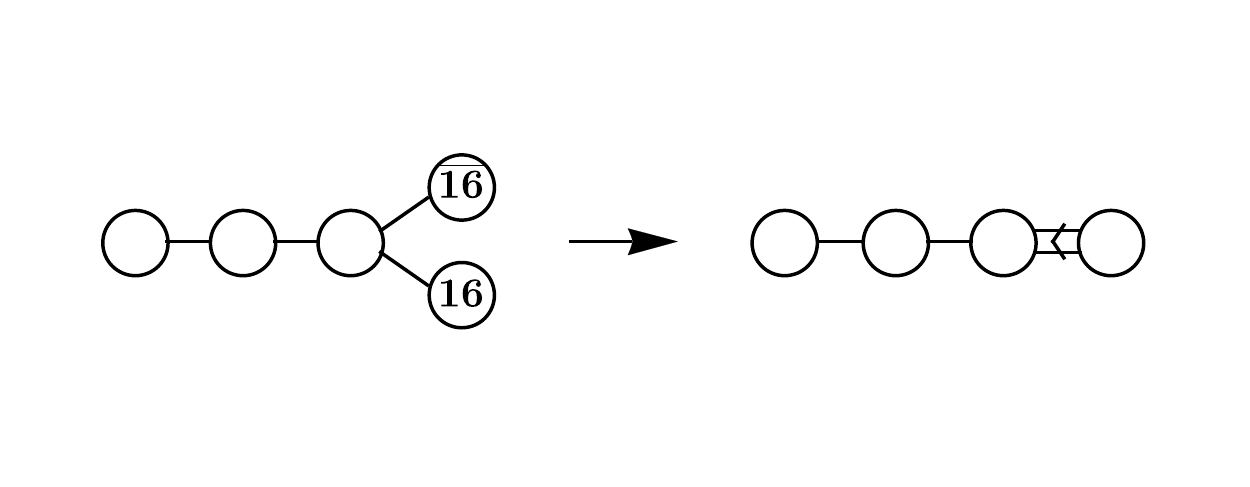}
  \caption{The outer-automorphism acting on $D_5$ diagram of $\mathfrak{so}(10)$ leads to $B_4$ diagram of $\mathfrak{so}(9)$.}
  \label{fig:I1*_dyn_mono}
\end{figure*}
This gives the expected reduction from $SO(10)$ to $SO(9)$.

\subsubsection{Type $I_4$: $SU(4)\rightarrow Sp(2)$}\label{sec:I4analytic}

The geometric monodromy action was analyzed in Section \ref{sec:I4_geom_analytic}. The ordered set of seven branes at $z = 0$ is $\{1,1,1,1\}$. The intersection matrix is:
\begin{center}
  $I = 
  \begin{pmatrix}
    -1 & 0 & 0 & 0 \\
    0 & -1 & 0 & 0 \\
    0 & 0 & -1 & 0 \\
    0 & 0 & 0 & -1
  \end{pmatrix}$.
\end{center}
The simple roots in junction basis are:
\begin{align*}
  \alpha_1 &= (0, 0, 1, -1) \\
  \alpha_2 &= (0, 1, -1, 0) \\
  \alpha_3 &= (1, -1, 0, 0).
\end{align*}
The Cartan matrix is:
\[
\begin{pmatrix}
    -2 & 1 & 0 \\
    1 & -2 & 1 \\
    0 & 1 & -2 
\end{pmatrix}.
\]
and the junction-to-Dynkin map is:
\[
T
= 
\begin{pmatrix}
    0 & 0 & 1 & -1 \\
    0 & 1 & -1 & 0 \\
    1 & -1 & 0 & 0
\end{pmatrix}.
\]

The highest weight junctions of various representations and asymptotic charges that will be important for us are:
\begin{center}
\begin{tabular}{c|c|c|c}
Asymptotic Charge & Highest Weight Junction & Dynkin Label & Representation\\
\hline
$(1,0)$ & $(1,0,0,0)$ & $[0,0,1]$ & $\mathbf{4}$ \\
\hline
$(-1,0)$ & $(0,0,0,-1)$ & $[1,0,0]$ & $\mathbf{\bar{4}}$
\end{tabular}
\end{center}
We recognize that $[0,0,1]$ as the highest weight state of $\mathbf{4}$ and $[1,0,0]$ as the highest weight state of $\mathbf{\bar{4}}$. See Appendix \ref{sec:I4_junc_comp} for further details.

In this case the geometric monodromy action an $x_1$-$x_3$ swap, together with an overall $2\pi$ rotation (derived in Section~\ref{sec:I4_geom_analytic}). The 7-brane configuration in this case is $\{1, 1, 1, 1\}$, and therefore the $SL(2, \mathbb{Z})$ matrix corresponding to a single 7-brane is $M_1$. We may therefore conclude that the total geometric monodromy action is $-M_1$.

The $I_1$ monodromy
\begin{equation}
	-M_1 = 
	\begin{pmatrix}
		-1 & -1 \\
		0 & -1
	\end{pmatrix}\, ,
\end{equation}
then induces a transformation on the $(1,0)$ asymptotic charge of the string junctions
\begin{equation}
	\begin{pmatrix}
		1 \\ 0
	\end{pmatrix}
	\qquad \mapsto \qquad
	\begin{pmatrix}
		-1 \\ 0
	\end{pmatrix}\, ,
\end{equation}
which corresponds to a map on representations
\begin{equation}
	\mathbf{4}\qquad\mapsto\qquad\mathbf{\bar{4}}\, .
\end{equation}
Traversing a loop enclosing the other group of roots induces another $-M_1$ action in the asymptotic charge
\begin{equation}
	\begin{pmatrix}
		-1 \\ 0
	\end{pmatrix}
	\qquad \mapsto \qquad
	\begin{pmatrix}
		1 \\ 0
	\end{pmatrix}
\end{equation}
which corresponds to a map on representations
\begin{equation}
	\mathbf{\bar{4}}\qquad\mapsto\qquad\mathbf{4}.
\end{equation}
Therefore $\mathbf{4}$ and $\mathbf{\bar{4}}$ should be identified which corresponds to a $\mathbb{Z}_2$ outer-automorphism acting on $A_3$ Dynkin diagram which leads to a $C_2$ Dynkin diagram via identifying the left-most node and the right-most node,
reducing the symmetry from $SU(4)$ to $Sp(2)$ as expected. 

\begin{figure*}[ht]
  \centering
  \includegraphics[width=.5\linewidth]{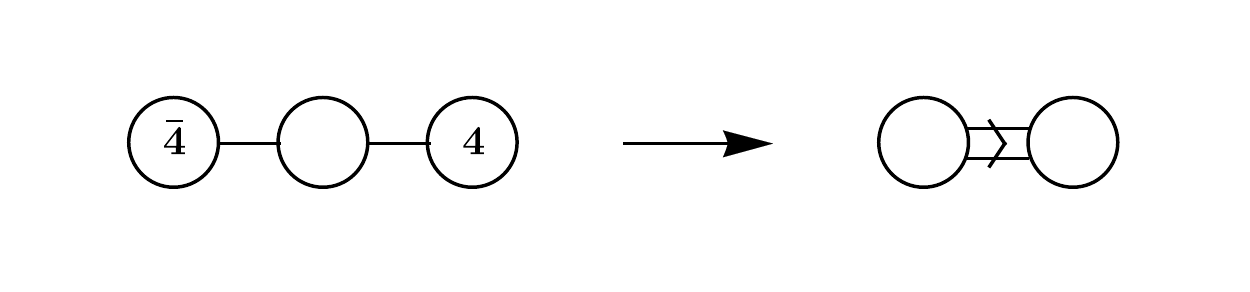}
  \caption{The outer-automorphism acting on $A_3$ diagram of $\mathfrak{su}(4)$ leads to $C_2$ diagram of $\mathfrak{sp}(2)$.}
  \label{fig:I4_dyn_mono}
\end{figure*}

\section{Discussion}
We have derived a classic result in F-theory, the Higgsing of simply-laced symmetry
algebras to ones that are non-simply-laced, in F-theory on a singular space. Previously
this result has been computed by resolving or deforming to a smooth space.

The origin of the effect is simple to understand. We considered one-parameter 
families of Weierstrass models with a non-abelian seven-brane on $D$, where
for $\epsilon=0$ the  symmetry algebra is simply-laced, but for $\epsilon\neq 0$
it is not. In all examples, the deformation to $\epsilon\neq 0$, which leaves the
variety singular, splits the $I_1$ locus ($\Delta_R=0$) within $D$, and
gives 
rise to new non-trivial loops in the geometry. Pulling the loop infinitesimally
away from $D$, we traverse it with a D3-brane. The monodromy associated with
the loop induces an action on $3$-$7$ string junctions that gives a non-trivial
map on flavor representations, signaling the reduction of the gauge algebra that arises
by quotienting by the associated outer-automorphism.

By treating the problem directly on the singular space, we were able to isolate
the feature critical for monodromy reduction: this splitting of the $I_1$ locus
inside the seven-brane. Performing the analysis on the singular space, however, required
motivating string junctions on the singular space. On a space smoothed by
deformation, the Lie algebraic
data associated to string junctions is derived from an ordered set of vanishing cycles,
a notion of asymptotic charge, and an appropriate pairing. The notion of asymptotic
charge is also natural on the singular space, and using a result from the math literature
we argued that (up to Hurwitz moves), a canonical ordered set of vanishing cycles is, as
well. We then took the pairing that is natural from string junctions, and performed
the analysis. We will motivate the pairing on the singular space and demonstrate
that it is well-behaved under Hurwitz moves in \cite{bcjt}.

Continued progress in understanding F-theory on singular spaces could be
of broad use, e.g., for the landscape, for its low energy effective supergravity theories,
and for its SCFT sectors. We plan on revisiting some of these issues in the future.

\vspace{.5cm} \noindent \textbf{Acknowledgments.} We thank Andres Collinucci, Ben Sung, and Roberto Valandro for useful discussions. We thank the Banff International Research Station for hospitality during part of this work. J.H. and C.L. are supported by NSF grant PHY-1620526.

\appendix

\section{String Junction Data}
\label{sec:junction_computations}

Here we will list the relevant information of the string junctions states in the junction basis and the junction-to-Dynkin maps that are needed for the results in the main text.

To verify our claims in Section \ref{sec:junc_analysis} we can, at zeroth order, check whether the number of string junctions with a given asymptotic charge matches the number of the states in the corresponding representations that we have specified. This counting is self-evident in the computations in this section and they indeed match. The first order check is to verify whether the spindle shaped structure of the states in a given representation is reproduced by a string junction computation. This also becomes obvious in our computation, and the reader can verify that such structure indeed appears. A final check would be to convert all information in the junction basis to Dynkin basis and check if the string junctions are indeed generated, and ordered in a manner such that the charges in Dynkin basis match the weights of the states in the claimed representation. We have checked that this is indeed the case.

Here we will list all the string junctions that are relevant in our discussion in Section \ref{sec:junc_analysis} and order them in a manner that both the number and the spindle shaped structure of the states are manifest. We will only present the highest weight states of the string junctions in Dynkin basis, in order to keep this appendix as concise as possible. We refer to~\cite{Grassi:2013kha} for a more in-depth discussion of the tools we utilize.

\subsection{$I_{0s}^*\rightarrow I_{0ns}^*$}\label{sec:G2_junc_comp}

We start with a $(1,0)$ string junction. Let us focus on one of the three splitting pairs of the roots of $\Delta_R(t; \delta; \epsilon) = 0$. The monodromy $M_3$ transforms it to a $(1,-1)$ string junction. The other $M_3$ action will then brings it to a $(1,-2)$ string junction. Focusing on the second pair, we see the monodromy $M_2$ brings it to a $(0,-1)$ string junction. The other $M_2$ action will then brings it to a $(-1,-2)$ string junction. If instead we start with a $(0,-1)$ string junction and encircle the second pair, the monodromy $M_1$ brings it to a $(-1,-1)$ string junction. The other $M_1$ action will then brings it back to a $(0,-1)$ string junction.

The $(1,0)$ string junctions in the junction basis are:
\begin{center}
\begin{tabular}{c|c|c}
Level & Mult. & Junctions \\ \hline
0&1&(1, 0, 0, 0, 0, 0)\,\,\, \\
1&1&(0, 0, 1, 0, 0, 0)\,\,\, \\
2&1&(0, 0, 0, 0, 1, 0)\,\,\, \\
3&2&(0, 0, 0, -1, 1, 1)\,\,\,(0, -1, 1, 1, 0, 0)\,\,\, \\
4&1&(0, -1, 1, 0, 0, 1)\,\,\, \\
5&1&(0, -1, 0, 0, 1, 1)\,\,\, \\
6&1&(-1, -1, 1, 0, 1, 1)\,\,\, \\
\end{tabular}
\end{center}

The $(1,-1)$ string junctions in the junction basis are:
\begin{center}
\begin{tabular}{c|c|c}
Level & Mult. & Junctions \\ \hline
0&1&(0, 0, 0, -1, 1, 0)\,\,\, \\
1&1&(0, -1, 1, 0, 0, 0)\,\,\, \\
2&1&(0, -1, 0, 0, 1, 0)\,\,\, \\
3&2&(0, -1, 0, -1, 1, 1)\,\,\,(-1, -1, 1, 0, 1, 0)\,\,\, \\
4&1&(-1, -1, 1, -1, 1, 1)\,\,\, \\
5&1&(-1, -1, 0, -1, 2, 1)\,\,\, \\
6&1&(-1, -2, 1, 0, 1, 1)\,\,\, \\
\end{tabular}
\end{center}

The $(1,-2)$ string junctions in the junction basis are:
\begin{center}
\begin{tabular}{c|c|c}
Level & Mult. & Junctions \\ \hline
0&1&(0, -1, 0, -1, 1, 0)\,\,\, \\
1&1&(-1, -1, 1, -1, 1, 0)\,\,\, \\
2&1&(-1, -1, 0, -1, 2, 0)\,\,\, \\
3&2&(-1, -1, 0, -2, 2, 1)\,\,\,(-1, -2, 1, 0, 1, 0)\,\,\, \\
4&1&(-1, -2, 1, -1, 1, 1)\,\,\, \\
5&1&(-1, -2, 0, -1, 2, 1)\,\,\, \\
6&1&(-2, -2, 1, -1, 2, 1)\,\,\, \\
\end{tabular}
\end{center}

The $(0,-1)$ string junctions in the junction basis are:
\begin{center}
\begin{tabular}{c|c|c}
Level & Mult. & Junctions \\ \hline
0&1&(0, 0, 0, 0, 0, -1)\,\,\, \\
1&1&(0, 0, 0, -1, 0, 0)\,\,\, \\
2&1&(0, 0, -1, -1, 1, 0)\,\,\, \\
3&2&(0, -1, 0, 0, 0, 0)\,\,\,(-1, 0, 0, -1, 1, 0)\,\,\, \\
4&1&(-1, -1, 1, 0, 0, 0)\,\,\, \\
5&1&(-1, -1, 0, 0, 1, 0)\,\,\, \\
6&1&(-1, -1, 0, -1, 1, 1)\,\,\, \\
\end{tabular}
\end{center}

The $(-1,-2)$ string junctions in the junction basis are:
\begin{center}
\begin{tabular}{c|c|c}
Level & Mult. & Junctions \\ \hline
0&1&(0, 0, -1, -1, 0, -1)\,\,\, \\
1&1&(-1, 0, 0, -1, 0, -1)\,\,\, \\
2&1&(-1, 0, -1, -1, 1, -1)\,\,\, \\
3&2&(-1, 0, -1, -2, 1, 0)\,\,\,(-1, -1, 0, 0, 0, -1)\,\,\, \\
4&1&(-1, -1, 0, -1, 0, 0)\,\,\, \\
5&1&(-1, -1, -1, -1, 1, 0)\,\,\, \\
6&1&(-2, -1, 0, -1, 1, 0)\,\,\, \\
\end{tabular}
\end{center}

The $(-1,-1)$ string junctions in the junction basis are:
\begin{center}
\begin{tabular}{c|c|c}
Level & Mult. & Junctions \\ \hline
0&1&(0, 1, -1, -1, 0, -1)\,\,\, \\
1&1&(0, 0, 0, 0, -1, -1)\,\,\, \\
2&1&(0, 0, -1, 0, 0, -1)\,\,\, \\
3&2&(0, 0, -1, -1, 0, 0)\,\,\,(-1, 0, 0, 0, 0, -1)\,\,\, \\
4&1&(-1, 0, 0, -1, 0, 0)\,\,\, \\
5&1&(-1, 0, -1, -1, 1, 0)\,\,\, \\
6&1&(-1, -1, 0, 0, 0, 0)\,\,\, \\
\end{tabular}
\end{center}

The junction-to-Dynkin map is:
\[
T
= 
\begin{pmatrix}
  0 & 0 & 0 & 1 & 1 & -1 \\
  0 & 0 & 1 & -1 & -1 & 0 \\
  0 & 1 & 0 & 0 & 1 & 0 \\
  1 & -1 & -1 & 0 & 0 & 0
\end{pmatrix}.
\]

The reader can compare the results here and the discussions in Section.\ref{sec:G2_junc_analytic}.

\subsection{$I_{0s}^*\rightarrow I_{0ss}^*$}\label{sec:SO7_junc_comp}

As we have discussed in Section \ref{sec:SO7_junc_analytic}, the relevant junctions are those with asymptotic charges $(1,0)$, $(0,-1)$ and $(-1,-2)$. We have demonstrated that these junctions give rise to $\mathbf{8_v}$, $\mathbf{8_c}$ and $\mathbf{8_v}$ of $SO(8)$, and so the claim we made in Section \ref{sec:SO7_junc_analytic} that $\mathbf{8_v}$ is identified with $\mathbf{8_c}$ is verified. In Section \ref{sec:SO7_junc_analytic} we also claimed that the set of $(1,1)$ string junctions corresponds to $\mathbf{8_s}$ of $SO(8)$. We will show this is true via the same method as before.

The $(1,1)$ string junctions in the junction basis are:
\begin{center}
\begin{tabular}{c|c|c}
Level & Mult. & Junctions \\ \hline
0&1&(1, 1, 0, 0, 0, 0)\,\,\, \\
1&1&(1, 0, 1, 1, -1, 0)\,\,\, \\
2&1&(1, 0, 0, 1, 0, 0)\,\,\, \\
3&2&(1, 0, 0, 0, 0, 1)\,\,\,(0, 0, 1, 1, 0, 0)\,\,\, \\
4&1&(0, 0, 1, 0, 0, 1)\,\,\, \\
5&1&(0, 0, 0, 0, 1, 1)\,\,\, \\
6&1&(0, -1, 1, 1, 0, 1)\,\,\, \\
\end{tabular}
\end{center}

The junction-to-Dynkin map is the same as in the previous section, and so one can check this set of junctions with charge $(1,1)$ indeed corresponds $\mathbf{8_s}$ of $SO(8)$ with the highest weight state $[0,0,1,0]$.

\subsection{$IV_s\rightarrow IV_{ns}$}\label{sec:IV_junc_comp}

Here we start with a $(1,0)$ string junction. The monodromy $M_1\cdot M_3$ brings it to a $(0,-1)$ string junction. The other $M_1\cdot M_3$ action will then brings it to a $(-1,-1)$ string junction.

The $(1,0)$ string junctions in the junction basis are:
\begin{center}
\begin{tabular}{c|c|c}
Level & Mult. & Junctions \\ \hline
0&1&(1, 0, 0, 0)\,\,\, \\
1&1&(0, 0, 1, 0)\,\,\, \\
2&1&(0, -1, 1, 1)\,\,\, \\
\end{tabular}
\end{center}

The $(0,-1)$ string junctions in the junction basis are:
\begin{center}
\begin{tabular}{c|c|c}
Level & Mult. & Junctions \\ \hline
0&1&(0, 0, 0, -1)\,\,\, \\
1&1&(0, -1, 0, 0)\,\,\, \\
2&1&(-1, -1, 1, 0)\,\,\, \\
\end{tabular}
\end{center}

The $(-1,-1)$ string junctions in the junction basis are:
\begin{center}
\begin{tabular}{c|c|c}
Level & Mult. & Junctions \\ \hline
0&1&(0, 0, -1, -1)\,\,\, \\
1&1&(-1, 0, 0, -1)\,\,\, \\
2&1&(-1, -1, 0, 0)\,\,\, \\
\end{tabular}
\end{center}

The junction-to-Dynkin map is:
\[
T
= 
\begin{pmatrix}
    0 & 1 & 1 & -1\\
    1 & -1 & -1 & 0 \\
\end{pmatrix}.
\]

In Dynkin basis we see that the highest weight junction with charge $(1,0)$ and $(-1,-1)$ is $\mathbf{3}:[0,1]$ and that with charge $(0,-1)$ is $\mathbf{\bar{3}}:[1,0]$.

\subsection{$IV_s^*\rightarrow IV_{ns}^*$}\label{sec:IV*_junc_comp}

Here we start with a $(1,0)$ string junction. The monodromy $M_1\cdot M_3$ brings it to a $(0,-1)$ string junction. The other $M_1\cdot M_3$ action will then brings it to a $(-1,-1)$ string junction.

The $(1,0)$ string junctions in the junction basis are:
\begin{center}
\begin{tabular}{c|c|c}
Level & Mult. & Junctions \\ \hline
0&1&(1, 0, 0, 0, 0, 0, 0, 0)\,\,\, \\
1&1&(0, 0, 1, 0, 0, 0, 0, 0)\,\,\, \\
2&1&(0, 0, 0, 0, 1, 0, 0, 0)\,\,\, \\
3&1&(0, 0, 0, 0, 0, 0, 1, 0)\,\,\, \\
4&2&(0, 0, 0, 0, 0, -1, 1, 1)\,\,\,(0, 0, 0, -1, 1, 1, 0, 0)\,\,\, \\
5&2&(0, 0, 0, -1, 1, 0, 0, 1)\,\,\,(0, -1, 1, 1, 0, 0, 0, 0)\,\,\, \\
6&2&(0, 0, 0, -1, 0, 0, 1, 1)\,\,\,(0, -1, 1, 0, 1, 1, -1, 0)\,\,\, \\
7&2&(0, 0, -1, -1, 1, 0, 1, 1)\,\,\,(0, -1, 1, 0, 0, 1, 0, 0)\,\,\, \\
8&3&(0, -1, 0, 0, 1, 1, 0, 0)\,\,\,(-1, 0, 0, -1, 1, 0, 1, 1)\,\,\,(0, -1, 1, 0, 0, 0, 0, 1)\,\,\, \\
9&2&(0, -1, 0, 0, 1, 0, 0, 1)\,\,\,(-1, -1, 1, 0, 1, 1, 0, 0)\,\,\, \\
10&2&(0, -1, 0, 0, 0, 0, 1, 1)\,\,\,(-1, -1, 1, 0, 1, 0, 0, 1)\,\,\, \\
11&2&(0, -1, 0, -1, 1, 1, 0, 1)\,\,\,(-1, -1, 1, 0, 0, 0, 1, 1)\,\,\, \\
12&2&(-1, -1, 1, -1, 1, 1, 0, 1)\,\,\,(-1, -1, 0, 0, 1, 0, 1, 1)\,\,\, \\
13&1&(-1, -1, 0, -1, 2, 1, 0, 1)\,\,\, \\
14&1&(-1, -1, 0, -1, 1, 1, 1, 1)\,\,\, \\
15&1&(-1, -1, 0, -1, 1, 0, 1, 2)\,\,\, \\
16&1&(-1, -2, 1, 0, 1, 1, 0, 1)\,\,\, \\
\end{tabular}
\end{center}

The $(0,-1)$ string junctions in the junction basis are:
\begin{center}
\begin{tabular}{c|c|c}
Level & Mult. & Junctions \\ \hline
0&1&(0, 1, -1, -1, 0, -1, 1, 0)\,\,\, \\
1&1&(0, 0, 0, 0, 0, 0, 0, -1)\,\,\, \\
2&1&(0, 0, 0, 0, 0, -1, 0, 0)\,\,\, \\
3&1&(0, 0, 0, 0, -1, -1, 1, 0)\,\,\, \\
4&2&(0, 0, 0, -1, 0, 0, 0, 0)\,\,\,(0, 0, -1, 0, 0, -1, 1, 0)\,\,\, \\
5&2&(0, 0, -1, -1, 1, 0, 0, 0)\,\,\,(-1, 0, 0, 0, 0, -1, 1, 0)\,\,\, \\
6&2&(0, 0, -1, -1, 0, 0, 1, 0)\,\,\,(-1, 0, 0, -1, 1, 0, 0, 0)\,\,\, \\
7&2&(0, 0, -1, -1, 0, -1, 1, 1)\,\,\,(-1, 0, 0, -1, 0, 0, 1, 0)\,\,\, \\
8&3&(0, -1, 0, 0, 0, 0, 0, 0)\,\,\,(-1, 0, 0, -1, 0, -1, 1, 1)\,\,\,(-1, 0, -1, -1, 1, 0, 1, 0)\,\,\, \\
9&2&(-1, -1, 1, 0, 0, 0, 0, 0)\,\,\,(-1, 0, -1, -1, 1, -1, 1, 1)\,\,\, \\
10&2&(-1, -1, 0, 0, 1, 0, 0, 0)\,\,\,(-1, 0, -1, -1, 0, -1, 2, 1)\,\,\, \\
11&2&(-1, -1, 0, 0, 0, 0, 1, 0)\,\,\,(-1, 0, -1, -2, 1, 0, 1, 1)\,\,\, \\
12&2&(-1, -1, 0, 0, 0, -1, 1, 1)\,\,\,(-1, -1, 0, -1, 1, 1, 0, 0)\,\,\, \\
13&1&(-1, -1, 0, -1, 1, 0, 0, 1)\,\,\, \\
14&1&(-1, -1, 0, -1, 0, 0, 1, 1)\,\,\, \\
15&1&(-1, -1, -1, -1, 1, 0, 1, 1)\,\,\, \\
16&1&(-2, -1, 0, -1, 1, 0, 1, 1)\,\,\, \\
\end{tabular}
\end{center}

The $(-1,-1)$ string junctions in the junction basis are:
\begin{center}
\begin{tabular}{c|c|c}
Level & Mult. & Junctions \\ \hline
0&1&(1, 1, -1, 0, -1, -1, 0, -1)\,\,\, \\
1&1&(0, 1, 0, 0, -1, -1, 0, -1)\,\,\, \\
2&1&(0, 1, -1, 0, 0, -1, 0, -1)\,\,\, \\
3&1&(0, 1, -1, 0, -1, -1, 1, -1)\,\,\, \\
4&2&(0, 1, -1, 0, -1, -2, 1, 0)\,\,\,(0, 1, -1, -1, 0, 0, 0, -1)\,\,\, \\
5&2&(0, 1, -1, -1, 0, -1, 0, 0)\,\,\,(0, 0, 0, 1, -1, -1, 0, -1)\,\,\, \\
6&2&(0, 1, -1, -1, -1, -1, 1, 0)\,\,\,(0, 0, 0, 0, 0, 0, -1, -1)\,\,\, \\
7&2&(0, 1, -2, -1, 0, -1, 1, 0)\,\,\,(0, 0, 0, 0, -1, 0, 0, -1)\,\,\, \\
8&3&(0, 0, -1, 0, 0, 0, 0, -1)\,\,\,(-1, 1, -1, -1, 0, -1, 1, 0)\,\,\,(0, 0, 0, 0, -1, -1, 0, 0)\,\,\, \\
9&2&(0, 0, -1, 0, 0, -1, 0, 0)\,\,\,(-1, 0, 0, 0, 0, 0, 0, -1)\,\,\, \\
10&2&(0, 0, -1, 0, -1, -1, 1, 0)\,\,\,(-1, 0, 0, 0, 0, -1, 0, 0)\,\,\, \\
11&2&(0, 0, -1, -1, 0, 0, 0, 0)\,\,\,(-1, 0, 0, 0, -1, -1, 1, 0)\,\,\, \\
12&2&(-1, 0, 0, -1, 0, 0, 0, 0)\,\,\,(-1, 0, -1, 0, 0, -1, 1, 0)\,\,\, \\
13&1&(-1, 0, -1, -1, 1, 0, 0, 0)\,\,\, \\
14&1&(-1, 0, -1, -1, 0, 0, 1, 0)\,\,\, \\
15&1&(-1, 0, -1, -1, 0, -1, 1, 1)\,\,\, \\
16&1&(-1, -1, 0, 0, 0, 0, 0, 0)\,\,\, \\
\end{tabular}
\end{center}

The junction-to-Dynkin map is:
\[
T
= 
\begin{pmatrix}
    0 & 0 & 0 & 0 & 0 & 1 & 1 & -1 \\
    0 & 0 & 0 & 0 & 1 & -1 & -1 & 0 \\
    0 & 0 & 0 & 1 & 0 & 0 & 1 & 0 \\
    0 & 0 & 1 & -1 & -1 & 0 & 0 & 0 \\
    0 & 1 & 0 & 0 & 0 & 0 & 0 & 1 \\
    1 & -1 & -1 & 0 & 0 & 0 & 0 & 0
\end{pmatrix}.
\]

In Dynkin basis we see that the highest weight junction with charge $(1,0)$ and $(-1,-1)$ is $\mathbf{27}:[0,0,0,0,0,1]$ and that with charge $(0,-1)$ is $\mathbf{\overline{27}}:[0,0,0,0,1,0]$.

\subsection{$I_{1s}^*\rightarrow I_{1ns}^*$}\label{sec:I1*_junc_comp}

Here we start with a $(1,1)$ string junction state. The monodromy $M_1$ brings it to a $(2,1)$ string junction. The other $M_1$ action will then brings it to a $(3,1)$ string junction.

The $(1,1)$ string junctions in the junction basis are:
\begin{center}
\begin{tabular}{c|c|c}
Level & Mult. & Junctions \\ \hline
0&1&(1, 1, 0, 0, 0, 0, 0)\,\,\, \\
1&1&(1, 0, 1, 1, 0, 0, -1)\,\,\, \\
2&1&(1, 0, 1, 1, -1, 0, 0)\,\,\, \\
3&2&(1, 0, 1, 0, 0, 1, -1)\,\,\,(1, 0, 0, 1, 0, 0, 0)\,\,\, \\
4&2&(1, 0, 0, 0, 1, 1, -1)\,\,\,(0, 0, 1, 1, 0, 0, 0)\,\,\, \\
5&2&(1, 0, 0, 0, 0, 1, 0)\,\,\,(0, 0, 1, 0, 1, 1, -1)\,\,\, \\
6&2&(1, -1, 1, 1, 0, 1, -1)\,\,\,(0, 0, 1, 0, 0, 1, 0)\,\,\, \\
7&2&(0, -1, 2, 1, 0, 1, -1)\,\,\,(0, 0, 0, 0, 1, 1, 0)\,\,\, \\
8&1&(0, -1, 1, 1, 1, 1, -1)\,\,\, \\
9&1&(0, -1, 1, 1, 0, 1, 0)\,\,\, \\
10&1&(0, -1, 1, 0, 1, 2, -1)\,\,\, \\
\end{tabular}
\end{center}

The $(2,1)$ string junctions in the junction basis are:
\begin{center}
\begin{tabular}{c|c|c}
Level & Mult. & Junctions \\ \hline
0&1&(1, 0, 1, 1, 0, 0, 0)\,\,\, \\
1&1&(1, 0, 1, 0, 1, 1, -1)\,\,\, \\
2&1&(1, 0, 1, 0, 0, 1, 0)\,\,\, \\
3&2&(1, 0, 0, 0, 1, 1, 0)\,\,\,(1, -1, 2, 1, 0, 1, -1)\,\,\, \\
4&2&(1, -1, 1, 1, 1, 1, -1)\,\,\,(0, 0, 1, 0, 1, 1, 0)\,\,\, \\
5&2&(1, -1, 1, 1, 0, 1, 0)\,\,\,(0, -1, 2, 1, 1, 1, -1)\,\,\, \\
6&2&(1, -1, 1, 0, 1, 2, -1)\,\,\,(0, -1, 2, 1, 0, 1, 0)\,\,\, \\
7&2&(0, -1, 2, 0, 1, 2, -1)\,\,\,(0, -1, 1, 1, 1, 1, 0)\,\,\, \\
8&1&(0, -1, 1, 0, 2, 2, -1)\,\,\, \\
9&1&(0, -1, 1, 0, 1, 2, 0)\,\,\, \\
10&1&(0, -2, 2, 1, 1, 2, -1)\,\,\, \\
\end{tabular}
\end{center}

The $(3,1)$ string junctions in the junction basis are:
\begin{center}
\begin{tabular}{c|c|c}
Level & Mult. & Junctions \\ \hline
0&1&(1, 0, 1, 0, 1, 1, 0)\,\,\, \\
1&1&(1, -1, 2, 1, 1, 1, -1)\,\,\, \\
2&1&(1, -1, 2, 1, 0, 1, 0)\,\,\, \\
3&2&(1, -1, 2, 0, 1, 2, -1)\,\,\,(1, -1, 1, 1, 1, 1, 0)\,\,\, \\
4&2&(1, -1, 1, 0, 2, 2, -1)\,\,\,(0, -1, 2, 1, 1, 1, 0)\,\,\, \\
5&2&(1, -1, 1, 0, 1, 2, 0)\,\,\,(0, -1, 2, 0, 2, 2, -1)\,\,\, \\
6&2&(1, -2, 2, 1, 1, 2, -1)\,\,\,(0, -1, 2, 0, 1, 2, 0)\,\,\, \\
7&2&(0, -2, 3, 1, 1, 2, -1)\,\,\,(0, -1, 1, 0, 2, 2, 0)\,\,\, \\
8&1&(0, -2, 2, 1, 2, 2, -1)\,\,\, \\
9&1&(0, -2, 2, 1, 1, 2, 0)\,\,\, \\
10&1&(0, -2, 2, 0, 2, 3, -1)\,\,\, \\
\end{tabular}
\end{center}

The junction-to-Dynkin map is:
\[
T
= 
\begin{pmatrix}
  0 & 0 & 0 & 0 & 1 & -1 & -1 \\
  0 & 0 & 0 & 1 & 0 & 0 & 1 \\
  0 & 0 & 1 & -1 & -1 & 0 & 0 \\
  0 & 1 & 0 & 0 & 0 & 1 & 1 \\
  1 & -1 & -1 & 0 & 0 & 0 & 0
\end{pmatrix}.
\]
In Dynkin basis we see that the highest weight junction with charge $(1,1)$ is $\mathbf{16}:[0,0,0,1,0]$, that with charge $(2,1)$ is $\mathbf{\overline{16}}:[0,1,0,0,0]$ and that with charge $(3,1)$ is again $\mathbf{16}:[0,0,0,1,0]$. We have thus verified the results in Sec.~\ref{sec:I1*analytic}.

\subsection{$I_{4s}\rightarrow I_{4ns}$}\label{sec:I4_junc_comp}

Here we start with a $(1,0)$ string junction. The monodromy $-M_1$ brings it to a $(-1,0)$ string junction. The other $-M_1$ action will then brings it back to a $(1,0)$ string junction.

The $(1,0)$ string junctions in the junction basis are:
\begin{center}
\begin{tabular}{c|c|c}
Level & Mult. & Junctions \\ \hline
0&1&(1, 0, 0, 0)\,\,\, \\
1&1&(0, 1, 0, 0)\,\,\, \\
2&1&(0, 0, 1, 0)\,\,\, \\
3&1&(0, 0, 0, 1)\,\,\, \\
\end{tabular}
\end{center}
While $(-1,0)$ string junctions in the junction basis are:
\begin{center}
\begin{tabular}{c|c|c}
Level & Mult. & Junctions \\ \hline
0&1&(0, 0, 0, -1)\,\,\, \\
1&1&(0, 0, -1, 0)\,\,\, \\
2&1&(0, -1, 0, 0)\,\,\, \\
3&1&(-1, 0, 0, 0)\,\,\, \\
\end{tabular}
\end{center}
The junction-to-Dynkin map is:
\[
T
= 
\begin{pmatrix}
    0 & 0 & 1 & -1 \\
    0 & 1 & -1 & 0 \\
    1 & -1 & 0 & 0
\end{pmatrix}.
\]
In Dynkin basis we see that the highest weight junction with asymptotic charge $(1,0)$ is $\mathbf{4}:[0,0,1]$ and that with asymptotic charge $(-1,0)$ is $\mathbf{\bar{4}}:[1,0,0]$.

\section{Application of our method in type $II^*$ fibration}
\label{section:II*}

To further justify our method in Section \ref{sec:geom_monodromy}, we apply it to type $II^*$ fibration to obtain the expected seven brane configuration, giving rise to $E_8$ gauge group. The expected brane configuration is $\mathcal{S} = \{1,3,1,3,1,3,1,3,1,3\}$. Recall that we have discussed the correspondence between the relations between $U$ and $V$ realized at a generic point on the discriminant locus and and the type of the seven brane along the discriminant locus. In particular, we argued that $U_1$ corresponds to $1$ brane and $U_3$ corresponds to $3$ brane. In terms of the $U$-$V$ relations realized along the discriminant locus, we expect to see the alternating pattern $\{U_1, U_3, U_1, U_3, U_1, U_3, U_1, U_3, U_1, U_3\}$. Here we are only concerned with the alternating appearance of $U_1$ and $U_3$.

To separate and identify each seven brane in $\mathcal{S}$ and the $U$-$V$ relation along its locus we deform the Weierstrass model of type $II^*$:
\begin{equation}
  \begin{aligned}
    f &= f_1 z^4 + \epsilon, \\
    g &= g_1 z^5
  \end{aligned}
\end{equation}
where $\epsilon$ is the deformation parameter. This by no means is the most general form of deformation of the type $II^*$ Weierstrass model, but it will be enough for our purpose. Again we will keep $\epsilon$ small so that all the $I_1$ locus that we are interested in lie in a small neighborhood of certain point (which we consider to be the origin of a local patch) on the base manifold. In particular, we can treat sections of different line bundles over this local patch as complex functions.

The discriminant locus of out deformed type $II^*$ Weierstrass model is takes the form:
\begin{equation}
  4 f_1^3 z^{12}+12 f_1^2 z^8 \epsilon +12 f_1 z^4 \epsilon ^2+27 g_1^2 z^{10}+4 \epsilon ^3 = 0.
\end{equation}
There are twelve roots which parameterize the $I_1$ loci. One can show that there exist two roots $z_R = \pm \frac{3\sqrt{3}}{2}\sqrt{-\frac{g_1^2}{f_1^3}} + o(\epsilon^\frac{3}{10})$ out of the twelve roots that obviously do not approach the origin when $\epsilon\rightarrow 0$ whereas all the other ten roots are of order $O(\epsilon^\frac{3}{10})$, and so they collide at the origin when $\epsilon = 0$. It is then clear that it is the seven branes along these ten $I_1$ locus that form $\mathcal{S}$.

To the leading order of $\epsilon$, the ten roots that are relevant are:
\begin{equation}\label{eq:zsols_E8}
  \begin{aligned}
    z_R =& \  \bigg\{-\frac{\sqrt[5]{2} \sqrt{(-1)^{4/5} B_1}}{3^{3/10}}, \frac{\sqrt[5]{2}\sqrt{(-1)^{4/5} B_1}}{3^{3/10}}, -\frac{\sqrt[5]{2}\sqrt{B_1}}{3^{3/10}}, \frac{\sqrt[5]{2}\sqrt{B_1}}{3^{3/10}}, \\
      & \  -\frac{\sqrt[5]{2} \sqrt{-\sqrt[5]{-1}B_1}}{3^{3/10}},\frac{\sqrt[5]{2} \sqrt{-\sqrt[5]{-1}B_1}}{3^{3/10}},-\frac{\sqrt{(-2)^{2/5}B_1}}{3^{3/10}},\frac{\sqrt{(-2)^{2/5} B_1}}{3^{3/10}}, \\
      & \  -\frac{\sqrt[5]{2}\sqrt{-(-1)^{3/5} B_1}}{3^{3/10}},\frac{\sqrt[5]{2} \sqrt{-(-1)^{3/5}B_1}}{3^{3/10}}\bigg\}.
  \end{aligned}
\end{equation}
Here $B_1 = \epsilon^\frac{3}{5}(\frac{-1}{g_1^2})^\frac{1}{5}$.

We can still expand the solutions to $x^3 + fx + g = 0$ at one of the $I_1$ locus and the results are structurally similar to Eq.(\ref{eq:xsols_G2}):
\begin{equation}
  \begin{aligned}
    x_1 &= \frac{\sqrt[3]{\sqrt{A} W+V}}{3 \sqrt[3]{2}}-\frac{\sqrt[3]{2}U}{\sqrt[3]{\sqrt{A} W+V}}, \\
    x_2 &= \frac{\left(1+i \sqrt{3}\right) U}{2^{2/3}\sqrt[3]{\sqrt{A} W+V}}-\frac{\left(1-i \sqrt{3}\right) \sqrt[3]{\sqrt{A} W+V}}{6\sqrt[3]{2}}, \\
    x_3 &= \frac{\left(1-i \sqrt{3}\right) U}{2^{2/3} \sqrt[3]{\sqrt{A}W+V}}-\frac{\left(1+i \sqrt{3}\right) \sqrt[3]{\sqrt{A} W+V}}{6 \sqrt[3]{2}}.
  \end{aligned}
\end{equation}
Here $\sqrt{A}$ plays the role of $\sqrt{B_t}$ and of course $U$, $V$ and $W$ are different. Here $U = f_1 z^4 + \epsilon$ and $V = -27g_1 z^5$. In this case the relations between $U$ and $V$ are slightly modified to $108U^3 + V^2 = 0$ near the $I_1$ locus to the relevant order. Here we chose to modify the $U$-$V$ relation just for convenience, it is of no real significance.

We can now plug the values of $z_R$ into the expressions of $U$ and $V$ to see how the $x_i$-$x_j$ swaps are realized near each $z_R$. If the reader stares at Eq.(\ref{eq:zsols_E8}) long enough, they should recognize $z_R$'s are the tenth roots of unity besides the $\sqrt{B_1}$ factor and some multiplicative constant. Having observed this property, we can just set $B_1 = 1$ for simplicity since any non-zero value of $B_1$ will not affect the order of the roots. Changing the value of $B_1$ can be undone by a rotation of the reference coordinate system. 

The $I_1$ locus are shown in Figure.\ref{fig:check_E8}. Here we deliberately chose not to show the axis to remind the reader that although in order to illustrate the configuration of the roots we have chosen a particular value for $B_1$, the results do not depend on the chosen value, and different choices of $B_1$ are related by a rotation.
\begin{figure*}[ht]
  \centering
  \includegraphics[width=.3\linewidth]{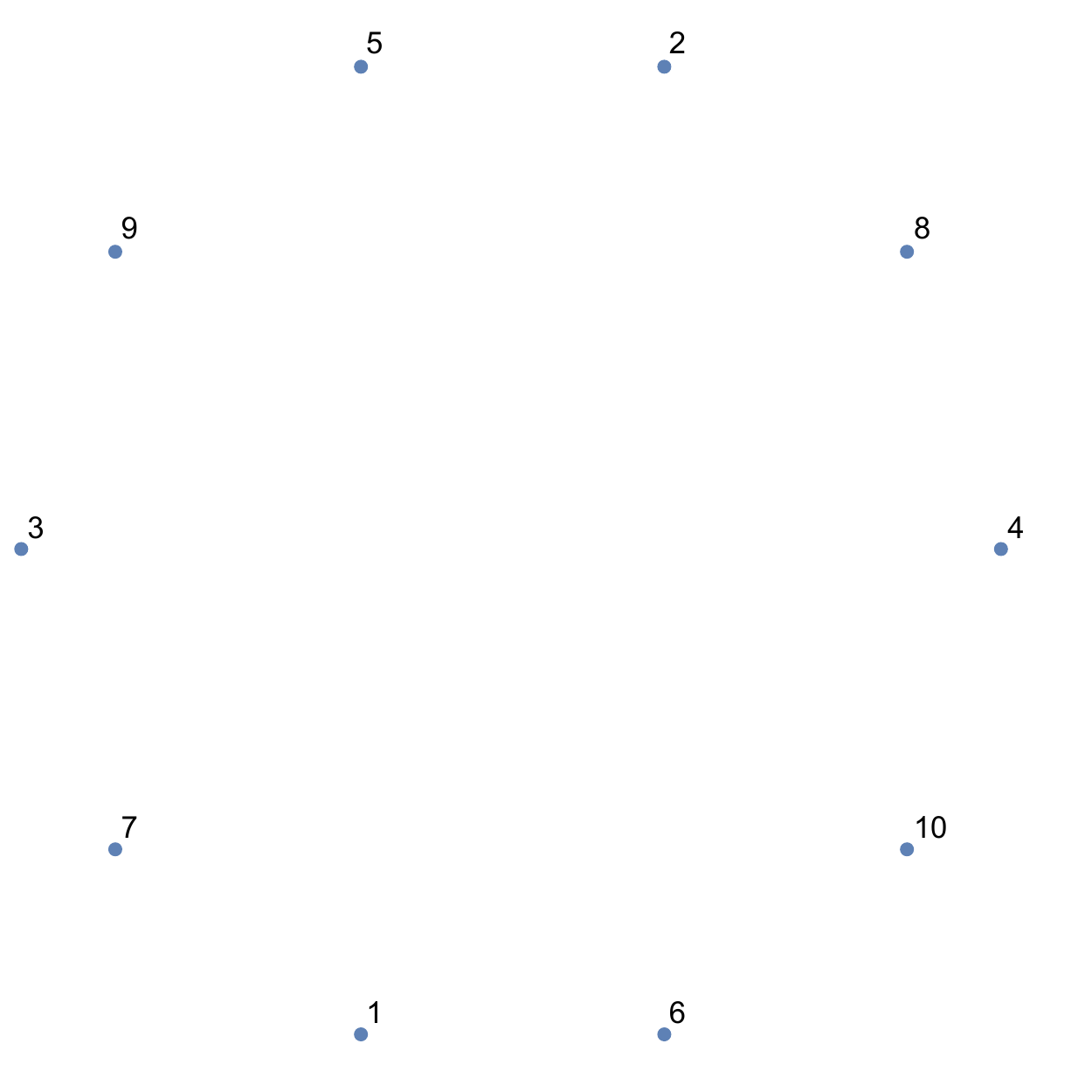}
  \caption{The $I_1$ locus of the deformed type $II^*$ model. The numbers labeling the points are the ordinals of the roots in Eq.(\ref{eq:zsols_E8}).}
  \label{fig:check_E8}
\end{figure*}

Applying the same method as in Section \ref{sec:cases_analytic} for obtaining the $U$-$V$ relation, it can be shown that the $U$-$V$ relations that are realized at the ten roots in Eq.(\ref{eq:zsols_E8}) are $U_1$, $U_3$, $U_1$, $U_3$, $U_1$, $U_3$, $U_3$, $U_1$, $U_3$, $U_1$, respectively.

Now it is easy see the pattern we are after, e.g., starting from the point labeled by 1 and traverse the roots in a clockwise manner, we see the alternating pattern:
\begin{align*}
  \{U_1, U_3, U_1, U_3, U_1, U_3, U_1, U_3, U_1, U_3\}.
\end{align*}

This completes our justification of the validity of our method for type $II^*$ fibration.

\bibliographystyle{JHEP}
\bibliography{ref_tian}
  
\end{document}